\documentclass{aastex701}
\usepackage{booktabs}
\usepackage{tablefootnote}
\usepackage{subfigure}
\usepackage{graphicx}
\usepackage[utf8]{inputenc}
\usepackage{CJK}
\usepackage{amsmath}
\usepackage{makecell}
\usepackage{caption}
\usepackage{natbib}
\begin{document}

\title{Distance Determination of Southern Galactic Plane Supernova Remnants with the Mopra CO Survey and DECaPS 3D Dust Map}

\begin{CJK*}{UTF8}{gbsn}

\correspondingauthor{He Zhao, Biwei Jiang}
\email{he.zhao@oca.eu, bjiang@bnu.edu.cn}

\author[0009-0007-3238-4212]{Fupeng Liu (刘付堋)}
\affiliation{School of Physics and Astronomy, Beijing Normal University, Beijing 100875, People's Republic of China}
\affiliation{Institute for Frontiers in Astronomy and Astrophysics, Beijing Normal University, Beijing 102206, People's Republic of China}
\email{202421101131@mail.bnu.edu.cn}

\author[0000-0003-2645-6869]{He Zhao (赵赫)}
\affiliation{Institute of Astronomy and Physics, Inner Mongolia University, Hohhot 010021, People's Republic of China}
\affiliation{Departamento de Fisica y Astronomia, Facultad de Ciencias Exactas, Universidad Andres Bello, Fernandez Concha 700, 8320000 Santiago, Chile}
\email{he.zhao@oca.eu}

\author[0000-0003-3168-2617]{Biwei Jiang (姜碧沩)}
\affiliation{School of Physics and Astronomy, Beijing Normal University, Beijing 100875, People's Republic of China}
\affiliation{Institute for Frontiers in Astronomy and Astrophysics, Beijing Normal University, Beijing 102206, People's Republic of China}
\email{bjiang@bnu.edu.cn}

\author[0000-0001-9328-4302]{Jun Li (李军)}
\affiliation{Center for Astrophysics, Guangzhou University, Guangzhou 510006, People's Republic of China}
\email{lijun@gzhu.edu.cn}

\author[0000-0002-1007-3700]{Zhe Zhang (张哲)}
\affiliation{School of Physics and Astronomy, Beijing Normal University, Beijing 100875, People's Republic of China}
\affiliation{Institute for Frontiers in Astronomy and Astrophysics, Beijing Normal University, Beijing 102206, People's Republic of China}
\email{zhangz324@mail.bnu.edu.cn}


\begin{abstract}

Accurate distance measurements to supernova remnants (SNRs) are crucial for understanding their physical properties, evolutionary processes, and role in the Galactic interstellar medium (ISM) cycle. In this study, we apply for the first time to the southern Galactic plane a distance determination method that utilizes CO emission data from the Mopra survey to identify molecular clouds (MCs) interacting with SNRs. By combining this with extinction--distance profiles from the DECaPS three-dimensional (3D) extinction map, we directly measure the distances to the associated MCs, thereby obtaining precise distances to the remnants. To overcome the extinction-missing bias in extremely dense regions where the 3D map suffers from a deficit of background stars, we supplement our analysis with two-dimensional (2D) extinction maps as cross-validation. Applying this method, we have derived precise distances for nine~SNRs: G290.1$-$0.8 ($7.32^{+0.60}_{-0.47}$~kpc), 
G292.2$-$0.5 ($10.85^{+0.43}_{-0.68}$~kpc), 
G296.1$-$0.5 ($4.59^{+0.18}_{-0.19}$~kpc), 
G296.8$-$0.3 ($8.74^{+0.40}_{-0.29}$~kpc), 
G298.6$-$0.0 ($6.50 \pm 0.21$~kpc), 
G312.4$-$0.4 ($3.60^{+0.19}_{-0.23}$~kpc), 
G332.4$-$0.4 ($2.66^{+0.23}_{-0.15}$~kpc), 
G335.2$+$0.1 ($2.76^{+0.37}_{-0.31}$~kpc), and 
G353.6$-$0.7 ($1.81^{+0.18}_{-0.14}$~kpc). Additionally, we established a robust lower distance limit of 1.34~kpc for G351.7$+$0.8.

\end{abstract}



\section{Introduction}

Supernova remnants (SNRs) play a crucial role in the evolution of the Milky Way. They are not only major sources of heavy elements injected into the interstellar medium (ISM), but also drive turbulence, heat the surrounding gas through shocks, and potentially trigger new star formation \citep{1977ApJ...214..725E,1995ApJS..101..181W}. Accurate distance measurements are essential for a detailed understanding of the physical properties of SNRs\textemdash including progenitor type, age, and feedback on the surrounding environment\textemdash because key quantities such as the true radius, luminosity, and expansion energy directly depend on the distance \citep{2018MNRAS.477.2243R}.

Currently, commonly used methods for estimating the distances to SNRs can be broadly classified into a few principal categories. The first is the kinematic method, which infers distances based on the Galactic rotation curve, using either radial velocities measured from HI absorption spectra or the velocities of molecular clouds (MCs) or HI structures that are associated with the remnant in both spatial and kinematic space \citep{2006MNRAS.369..416R, 2018AJ....155..204R, 2018MNRAS.477.2243R, 2022ApJ...940...63R}. This approach, however, is often limited by the near--far distance ambiguity, and its reliance on the rotation model, together with the presence of local noncircular motions in the Galaxy, can introduce significant uncertainties \citep{2009ApJ...699.1153R}.

The second is the empirical surface brightness-diameter ($\Sigma$--$D$) relation, which is widely used in a statistical sense \citep{1998ApJ...504..761C,2013ApJS..204....4P}. Due to the differences in the types of supernova explosions and the ISM environments in which they occur in the Milky Way, SNRs evolve in interstellar environments with varying properties. The efficiency of their radio emission mechanisms may be significantly affected, leading to considerable uncertainties in the estimation of the remnant's diameter, which could, in turn, introduce significant systematic errors \citep{2005MmSAI..76..534G}.

In recent years, extinction-based distance determination methods have gained increasing prominence. The core premise of this approach resides in the robust physical association or interaction between SNRs and their parental MCs. This is predicated on the fact that the majority of SNRs originate from core-collapse supernovae; their massive progenitor stars form within MC environments and, owing to their short evolutionary timescales, typically explode in close proximity to their birth sites. It is noteworthy that even Type~Ia SNRs may form in the vicinity of dense molecular gas \citep{2004ApJ...605L.113L,2016ApJ...833....4Z,2017A&A...604A..13C}. This physical proximity results in significantly elevated dust densities surrounding the remnants compared to the diffuse ISM, rendering the extinction jumps along the line of sight---arising from these associated MCs---reliable observational diagnostics for constraining the distances to the remnants. In practice, such extinction measurements can be performed using various celestial objects as tracers. For instance, \citet{2017MNRAS.472.3924C} and \citet{2019MNRAS.488.3129Y} employed three-dimensional (3D) extinction maps to infer distances via jump analysis, while others \citep[e.g.,][]{2018ApJ...855...12Z, 2020ApJ...891..137Z, 2025ApJ...988..176C} utilized spectroscopic information of background stars to reconstruct the line-of-sight extinction distribution. Additionally, \citet{2020A&A...639A..72W} adopted red clump stars as specific tracers to extract jump signals in the extinction profiles.

While this method shows promise in studies of individual SNRs, it faces fundamental limitations in practical application. Most SNRs are located in the dusty regions of the Galactic plane, where the line-of-sight extinction profile is often contaminated by unrelated MC structures. Relying solely on extinction jump signals identified by the aforementioned tracers, it is difficult to effectively distinguish whether the signal originates from dust associated with the SNR, foreground MCs, or background MCs. Moreover, existing methods typically detect only one or two prominent extinction jumps \citep{2020ApJ...891..137Z, 2025ApJ...988..176C}, which tend to be dominated by MCs with the highest column density or largest angular extent along the line of sight. However, such clouds may not necessarily be physically associated with the SNR. This ambiguity in source attribution introduces significant uncertainties in the measured distances, especially in regions of the Galactic disk where MCs are densely distributed.

To address the challenge of distinguishing MCs physically interacting with SNRs from unrelated line-of-sight MCs based solely on extinction jumps, \citet{2026arXiv260320881Z} recently developed a distance determination method based on the spatial correspondence between MCs and extinction structures, and successfully measured precise distances for four~SNRs. Their method first identifies the velocity range of MCs that are likely interacting with SNRs using prior association evidence. By slicing the 3D extinction maps, they correlate the spatial morphology of the CO integrated-intensity structures within this velocity range with the extinction structures in corresponding distance slices, thereby determining the distance of the associated MC as that of the SNR. This approach shifts the focus of distance determination from ambiguous extinction jump signals to MCs with well-defined spatial-kinematic correspondence, effectively mitigating contamination from unrelated line-of-sight structures.

In this work, we apply the same method for the first time to ten~SNRs located in the southern Galactic plane, using Mopra data \citep{2023PASA...40...47C, Burton2023}. A key extension of our study is that, in regions where the 3D extinction map suffers from insufficient background stars due to extremely high extinction, we additionally consult the 2D extinction map to identify observational biases (e.g., extinction-missing structures that arise from saturated opacity). This complementary use of 2D extinction information enhances the reliability of distance assignments in dense environments. Moreover, the high association rate ($\sim$80\%) between SNRs and MCs in the Milky Way, as revealed by the large-sample CO study of \citet{2023ApJS..268...61Z}, provides strong observational support for the broad applicability of this method.

This paper is organized as follows. Section~\ref{sec:data} describes the datasets employed in this study, including the CO survey data, 3D extinction maps, and the SNR sample selection. Section~\ref{sec:method} outlines our methodology for integrating these datasets to derive distances to the SNRs. Section~\ref{sec:RESULT AND DISCUSSION} presents the distance measurement results for the ten remnants and provides a comparative analysis with previous distance estimates. Finally, Section~\ref{sec:SUMMARY} summarizes our primary findings and conclusions.

\section{Data AND SNR SAMPLE}\label{sec:data} 

\subsection{Mopra CO survey} \label{}

The CO survey data utilized in this study are obtained from the Mopra CO Survey project\footnote{https://doi.org/10.25919/9z4p-mj92}, which was conducted between 2011 and 2018 using the 22\,m Mopra radio telescope in Australia \citep{2023PASA...40...47C, Burton2023}. Data Release 4 (DR4) of the survey was officially published in 2023, covering a Galactic longitude range of $250^\circ \le l \le 355^\circ$ and a latitude range of $|b| \le 1^\circ$, resulting in a total sky coverage exceeding $210\ \mathrm{deg^{2}}$. This study employs the $^{12}$CO (hereafter CO) $(J=1-0)$ emission line data from this survey, which features a spatial resolution of $0.6'$ and a velocity resolution of $0.1\ \mathrm{km\ s^{-1}}$.

\subsection{DECaPS 3D dust map} \label{}

We employ the DECaPS 3D dust extinction map published by \cite{2025ApJ...992...39Z}. This map represents the state-of-the-art 3D dust distribution model for the Galactic midplane. It is constructed from DECaPS2 survey data \citep{2023ApJS..264...28S}, integrated with VVV \citep{2010NewA...15..433M} and 2MASS near-infrared photometry \citep{2006AJ....131.1163S}, unWISE mid-infrared observations \citep{2019ApJS..240...30S}, and Gaia DR3 \citep{2023A&A...674A...1G} parallaxes. The model incorporates approximately 793 million stellar sources. While its total sample size is comparable to that of the Bayestar 3D dust map \citep{2019ApJ...887...93G} ($\sim 799$ million stars), its more focused sky coverage ($239^\circ < l < 6^\circ$, $-10^\circ < b < 10^\circ$) results in a significantly higher median stellar source density of 70 stars per arcmin$^2$. This substantially enhances the spatial resolution in the Galactic midplane, achieving an angular resolution of $1'$.

To ensure the robustness of our distance determinations, we strictly adhere to the reliability criteria defined by \cite{2025ApJ...992...39Z}. The study indicates a median maximum reliable distance of $\sim 9.4$ kpc (with a $1\sigma$ range of 7.6--12.2 kpc); all of our SNRs distance measurements fall within this empirically validated high-confidence range. The map employs a distance modulus binning of $\Delta \mu = 0.125$ mag, corresponding to a distance resolution of $\approx 6\% \cdot d$ (i.e., $\sim 0.12$ kpc at 2 kpc and $\sim 0.58$ kpc at 10 kpc), at greater distances ($d \gtrsim 8$ kpc), the actual distance uncertainty likely exceeds this theoretical binning limit due to the decline in stellar density and the accumulation of photometric errors within the DECaPS dataset. Overall, our distance determination precision is primarily constrained by the intrinsic distance modulus binning of the 3D extinction model and the statistical sampling density along the line of sight. Extinction data are retrieved via the Python package dustmaps\footnote{https://dustmaps.readthedocs.io/}.

In regions of high extinction where 3D maps struggle to provide accurate traces, we supplement our analysis with two-dimensional (2D) extinction maps for cross-verification. These include the high-resolution near-infrared extinction map from \hypertarget{ZK22_definition}{\citet[hereafter ZK22]{2022MNRAS.517.5180Z}} (spatial resolution $30''$) and the improved all-sky dust map from \hypertarget{CSFD_def}{\citet[hereafter CSFD]{2023ApJ...958..118C}} (spatial resolution $6'$).

\subsection{SNR Sample} \label{}

Due to the latitude coverage constraints ($|b| < 1^\circ$) of the Mopra CO survey, we selected ten SNRs to apply our proposed distance determination methodology: G290.1$-$0.8, G292.2$-$0.5, G296.1$-$0.5, G296.8$-$0.3, G298.6$-$0.0, G312.4$-$0.4, G332.4$-$0.4, G335.2$+$0.1, G351.7$+$0.8, and G353.6$-$0.7. The sample was chosen based on the following criteria: (1) the remnants possess relatively large angular sizes ($>10' \times 10'$), because the angular resolution limits of 3D extinction maps, larger remnants facilitate more reliable distance determination. (2) all targets are fully covered by both the Mopra CO survey and the DECaPS 3D extinction map; and (3) each remnant has established kinematic evidence (e.g., CO or HI velocity information) supporting its potential physical association with MCs. Other SNRs within the survey footprint were excluded primarily due to extreme angular sizes (too small or too large), severe source confusion along the line of sight, or a lack of reliable kinematic constraints. The physical parameters (obtained from SNRcat, \citealt{2012AdSpR..49.1313F}\footnote{\url{https://snrcat.physics.umanitoba.ca}}; and Green's Catalogue, \citealt{Green2024catalogue, 2025JApA...46...14G}\footnote{\url{https://www.mrao.cam.ac.uk/surveys/snrs/}}) and the velocity information of MCs potentially associated with the SNRs from previous studies are summarized in Table~\ref{tab1}.

The radio continuum data for the SNR sample in this study are drawn from two complementary sky surveys. For G290.1$-$0.8, G292.2$-$0.5, G296.1$-$0.5, G296.8$-$0.3, and G298.6$-$0.0, we preferentially used data from the Evolutionary Map of the Universe (EMU) survey at a frequency of 943~MHz \citep{2025PASA...42...71H}, which provides higher spatial resolution and more recent observations. For the remaining five remnants (G312.4$-$0.4, G332.4$-$0.4, G335.2$+$0.1, G351.7$+$0.8, and G353.6$-$0.7), which lie outside the current EMU coverage, we employed data from the Sydney University Molonglo Sky Survey (SUMSS) at 843~MHz \citep{2003MNRAS.342.1117M}. These radio continuum data are primarily utilized to precisely demarcate the morphological boundaries of the SNRs. Since the physical associations between the SNRs and MCs in this study are adopted from previous literature, we employ the radio contours to verify the spatial consistency between these known associations and the observed CO distribution.

\section{Method} \label{sec:method}

The distance determination in this work follows a systematic three-stage process, involving observational constraints, morphological matching, and gradient analysis of extinction profiles.

First, we perform observational calibration and velocity interval determination by identifying MC components associated with each SNR. 
Through detailed spectral analysis and spatial morphology matching of the Mopra CO survey data, and considering the differences in observational parameters compared to previous studies (Table~\ref{tab1}), we extract the average CO spectra from rectangular regions encompassing the radio continuum extent of each SNR. 
These regions are chosen to facilitate the consistent extraction of CO integrated-intensity maps and 3D differential extinction maps while ensuring strict spatial alignment across datasets. 
The RMS noise level is estimated from an emission-free velocity range of $200$--$400$\,km\,s$^{-1}$, incorporating baseline fluctuations to define a conservative $3\sigma_{\rm rms}$ threshold; only features with peak intensities above this threshold and spatially coherent structures in velocity-channel maps are considered genuine MC components. 
These features are compared with previously reported velocity ranges to verify consistency and refine velocity boundaries, allowing us to accurately identify MCs associated with each SNR and exclude foreground or background components. 
Based on the established velocity components, we then generate the corresponding velocity-integrated CO intensity maps.

Next, we conduct extinction slicing and morphological matching. 
Using the DECaPS 3D extinction map, we perform systematic slicing within 10 kpc with a coarse step of 1 kpc. 
This initial large-scale scanning efficiently identifies potential distance intervals where the extinction distribution morphologically matches the CO maps.

Finally, we analyze extinction-distance profiles to determine accurate distances and uncertainties. 
For each MC component, we select representative lines of sight (Table~\ref{tab2}) with prominent extinction jumps and extract $E(B-V)$ profiles from the DECaPS map. 
A smoothing spline is used to derive the differential gradient, representing the dust volume density along the line of sight. 
The distance is defined by the position of the gradient peak, and the uncertainty is taken as its full width at half maximum (FWHM), which reflects the physical thickness of the MC and the systematic distance uncertainty. 
For morphological verification, the distance interval at 10\% of the peak intensity serves as the integration range for the final differential extinction maps.

When MCs exhibit strong CO emission, their high internal dust density leads to very high extinction, causing a severe deficit of detectable background stars in the DECaPS 3D map (constructed via star counting) and degrading the distance accuracy. 
To address this limitation, we incorporate either the \hyperlink{CSFD_def}{CSFD} or \hyperlink{ZK22_definition}{ZK22} 2D extinction maps and use the CO integrated intensity to estimate the expected extinction. 
The expected visual extinction from the molecular gas is calculated via:
\begin{equation}
    A_V^{\text{expected}}  = \text{DGR} \times (2 \times X_{\text{CO}} \times W_{\text{CO}}),
\end{equation}
where $W_{\text{CO}} = \int T_{\mathrm{B}} \, \mathrm{d}v$ is the velocity-integrated $\mathrm{CO}$ intensity, and $(2 \times X_{\text{CO}} \times W_{\text{CO}})$ gives the total hydrogen column density (assuming purely molecular gas). 
We adopt the Galactic average dust-to-gas ratio $\text{DGR} = 5.34 \times 10^{-22}~\mathrm{mag~cm^2}$ \citep{1977ApJ...216..291S,1978ApJ...224..132B} and the standard $\mathrm{CO}$-to-$\mathrm{H}_2$ conversion factor $X_{\text{CO}} = 2 \times 10^{20}~\mathrm{cm^{-2}~(K~km~s^{-1})^{-1}}$ \citep{1982ApJ...262..590F,2001ApJ...547..792D,2006A&A...454..781L}. 

To facilitate comparison with the DECaPS 3D dust map, this value is converted to color excess via $E(B-V) = A_V^{\text{expected}}  / 3.1$, assuming a standard reddening law with $R_V = 3.1$. 
This expected extinction serves as a secondary check on the correspondence between molecular gas and dust; by comparing $A_V^{\text{expected}}$ with the observed jump $\Delta E(B-V)_{\rm obs}$ at the identified distance, we assess whether the detected MC component is the primary contributor to the extinction feature. 
Discrepancies are expected and likely reflect physical complexities such as variations in the $X_{\text{CO}}$ factor due to shock heating, the presence of CO-dark gas, or additional contributions from HI.

\section{RESULT AND DISCUSSION} \label{sec:RESULT AND DISCUSSION}
The CO integrated intensity maps covering all detected MC 
components within the field of each SNR are presented in Figures~\ref{fig:A1} and~\ref{fig:A2}, where the radio continuum contours of the remnants and their surrounding environments are selectively overlaid on the MC components 
that are potentially physically associated with the SNRs to highlight their spatial relationships. Among these, the specific MC components detectable with the DECaPS 3D extinction map, along with their corresponding distances, 
are listed in Table~\ref{tab2}. Other components not included in the table are either beyond the 10~kpc detection range or cannot be reliably extracted due to strong contamination from foreground MCs. For MCs confirmed to be potentially physically associated with the SNRs, the distances measured by this method are directly adopted as the distance estimates for the corresponding remnants. The finalized distance results for all sample SNRs are summarized in Table~\ref{tab3}.

\subsection{G290.1$-$0.8} \label{sec:4.1}
\begin{figure}[htb!]
   \centering
   \includegraphics[width=0.9\textwidth]{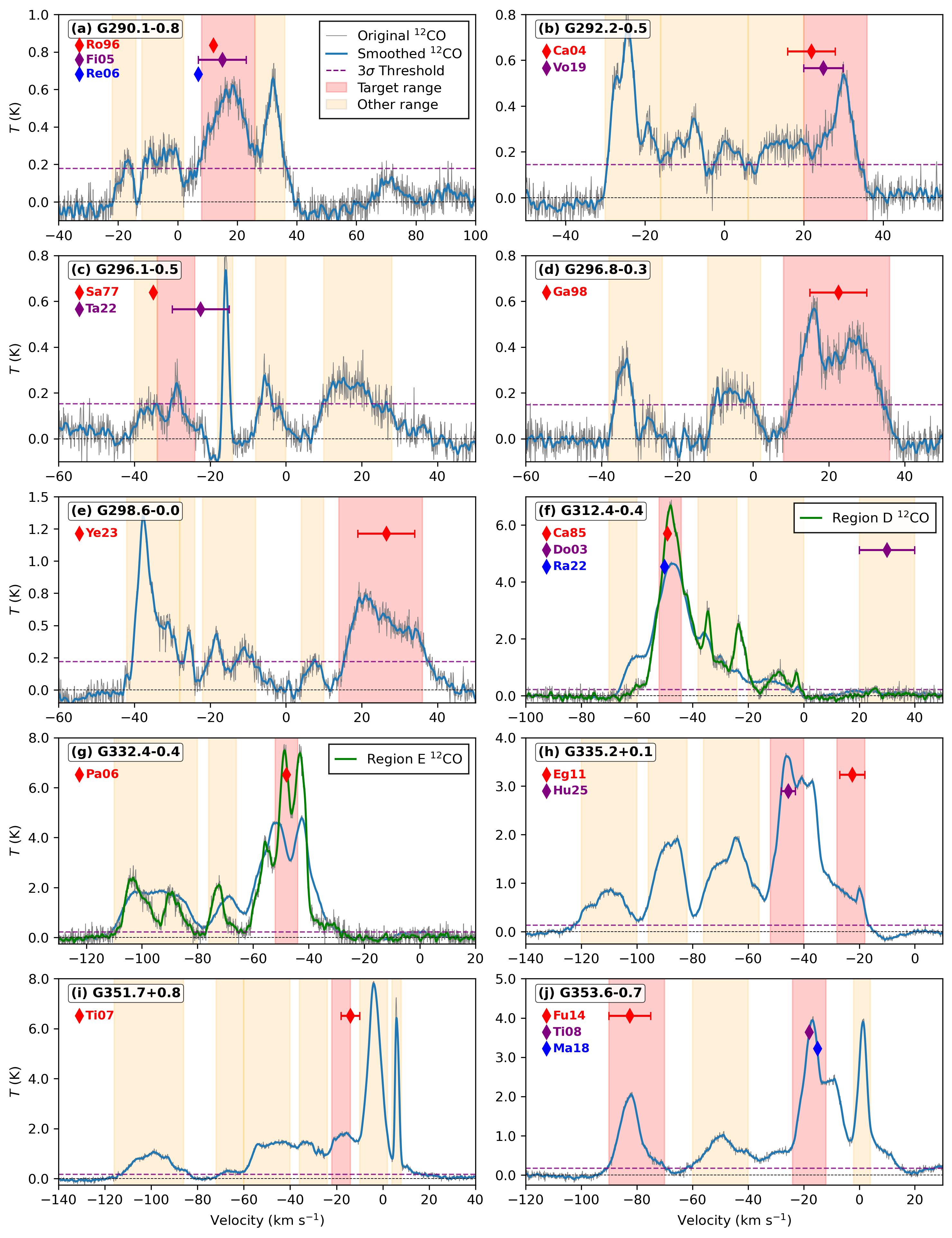} 
   \caption{Spatially averaged CO ($J = 1-0$) spectra extracted from the rectangular regions encompassing 10 SNRs: 
(a) G290.1$-$0.8, (b) G292.2$-$0.5, (c) G296.1$-$0.5, (d) G296.8$-$0.3, (e) G298.6$-$0.0, 
(f) G312.4$-$0.4, (g) G332.4$-$0.4, (h) G335.2$+$0.1, (i) G351.7$+$0.8, and (j) G353.6$-$0.7. 
In each panel, the gray curve represents the original spectrum, while the blue curve shows the smoothed profile. 
The purple dashed line indicates the $3\sigma$ noise level. 
The red shaded velocity intervals mark the CO velocity ranges of MCs that are likely physically interacting with each SNR, while the orange shaded regions indicate other, non-interacting MC components along the same line of sight. 
In panels (f) and (g), additional green curves display the spatially averaged CO spectra extracted from regions D and E (marked in Figure~\ref{fig:A2}), which aid in better tracing the MC components potentially interacting with the SNR. 
The colored diamonds with vertical bars indicate the proposed associated velocity ranges in the literature; the abbreviated reference labels (e.g., ''Ro96'' for \citealt{1996A&A...315..243R}) correspond to the full citations listed in Table~\ref{tab1}.} 
   \label{fig 1}
\end{figure}

G290.1$-$0.8 is a mixed-morphology SNR with a very complex ISM environment along the line of sight, containing multiple foreground and background MCs \citep{1996A&A...315..243R}. Based on H$\alpha$ emission-line observations, \citet{1996A&A...315..243R} inferred a radial velocity of approximately 12 km s$^{-1}$ for this remnant. \citet{2005SerAJ.170...47F} further suggested, through CO molecular line observations, that a MC located southwest of the remnant, with a velocity range of $[7, 23]$ km s$^{-1}$, might be physically associated with the remnant. \citet{2006MNRAS.369..416R} used HI absorption spectra to determine a local standard of rest (LSR) velocity of about 7 km s$^{-1}$ for this remnant. Based on velocity references from previous literature and the analysis of our mean CO spectra, we identified four primary MC components within the target region using a $3\sigma$ threshold (see Figures \ref{fig 1}a and \ref{fig:A1}). Their velocity ranges are $[-22, -14]$, $[-12, 2]$, $[8, 26]$, and $[26, 36]$~km~s$^{-1}$. Notably, the $[8, 26]$~km~s$^{-1}$ component exhibits intense CO emission to the southwest of the SNR, consistent with the findings of \citet{2005SerAJ.170...47F}. Consequently, we consider this component the most likely candidate for physical interaction with the SNR.

\begin{figure*}[htb!]
    \centering
    \includegraphics[width=1.0\textwidth]{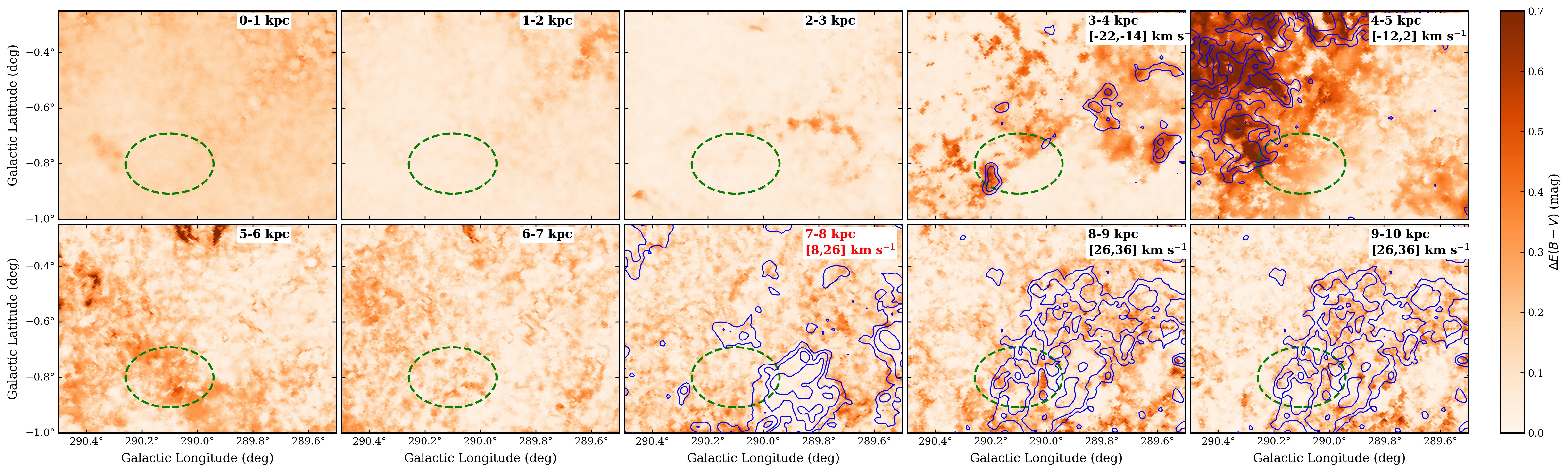}
    \caption{Uniform differential extinction maps for G290.1$-$0.8 at 1-kpc intervals within 10~kpc. Blue contours overlaid on the respective panels represent the CO emission from the velocity components associated with each distance interval. The dark green ellipse denotes the approximate boundary of the SNR. The red-labeled velocity and distance ranges indicate the molecular components and their corresponding distance estimates identified as being physically associated with the SNR.}
    \label{fig 2}
\end{figure*}

\begin{figure}[htb!]
    \centering
    \includegraphics[width=0.9\textwidth]{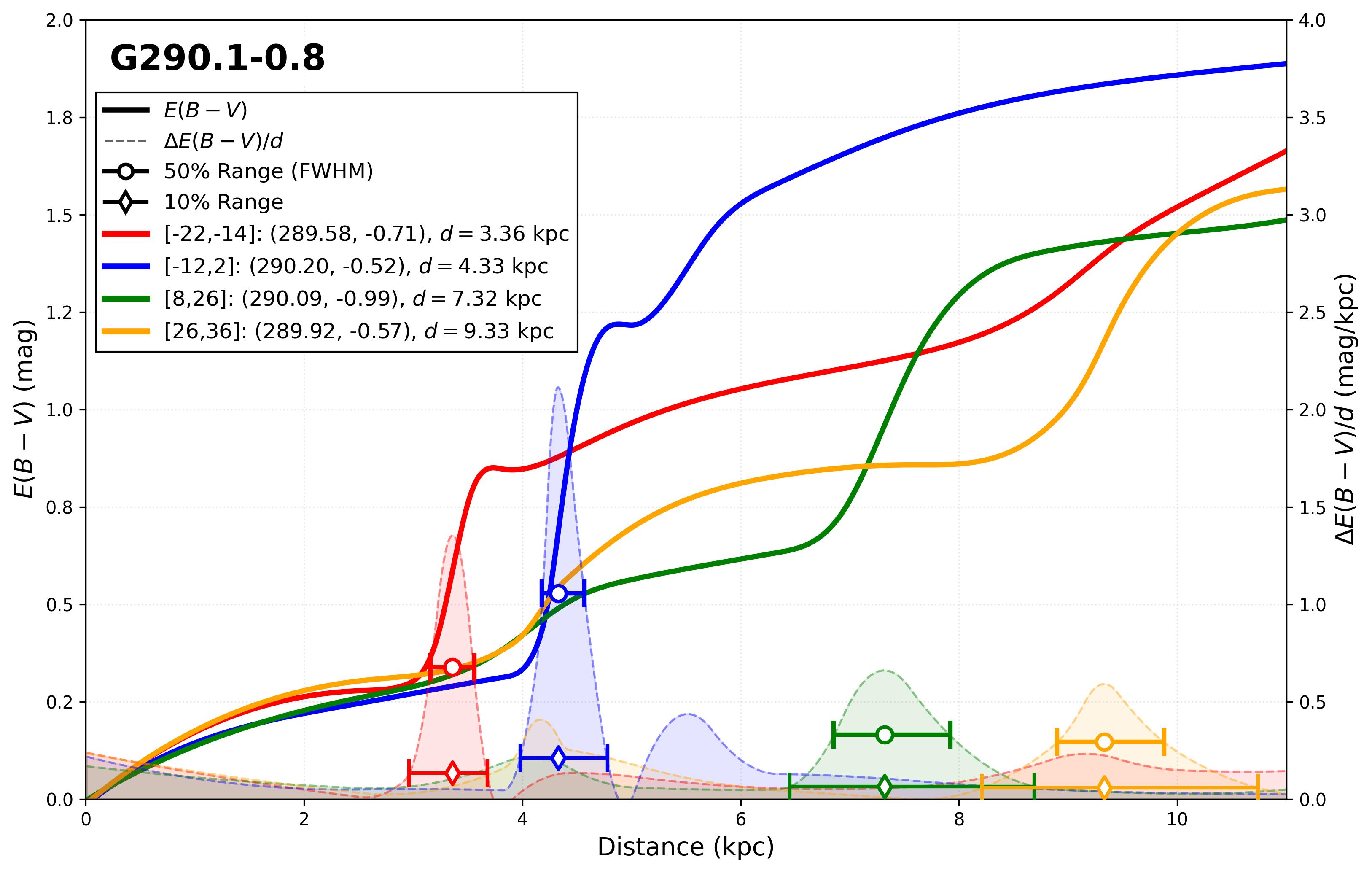}  
    \caption{Cumulative color excess $E(B-V)$ (solid lines) and extinction gradient (dashed lines) as a function of distance toward representative sightlines of different MC components for G290.1$-$0.8. The selected Galactic coordinates $(l, b)$ and the derived distances for each component are labeled in the respective panels, with their corresponding positions marked by red triangles in Figure~\ref{fig:A1}. For each MC component, the distance is determined by the gradient peak, and the uncertainty is defined by its FWHM. The distance interval at 10\% of the peak intensity is adopted as the integration thickness for the morphological verification.}  
    \label{fig 3}  
\end{figure}

\begin{figure}[htb!]
    \centering
    \includegraphics[width=0.9\textwidth]{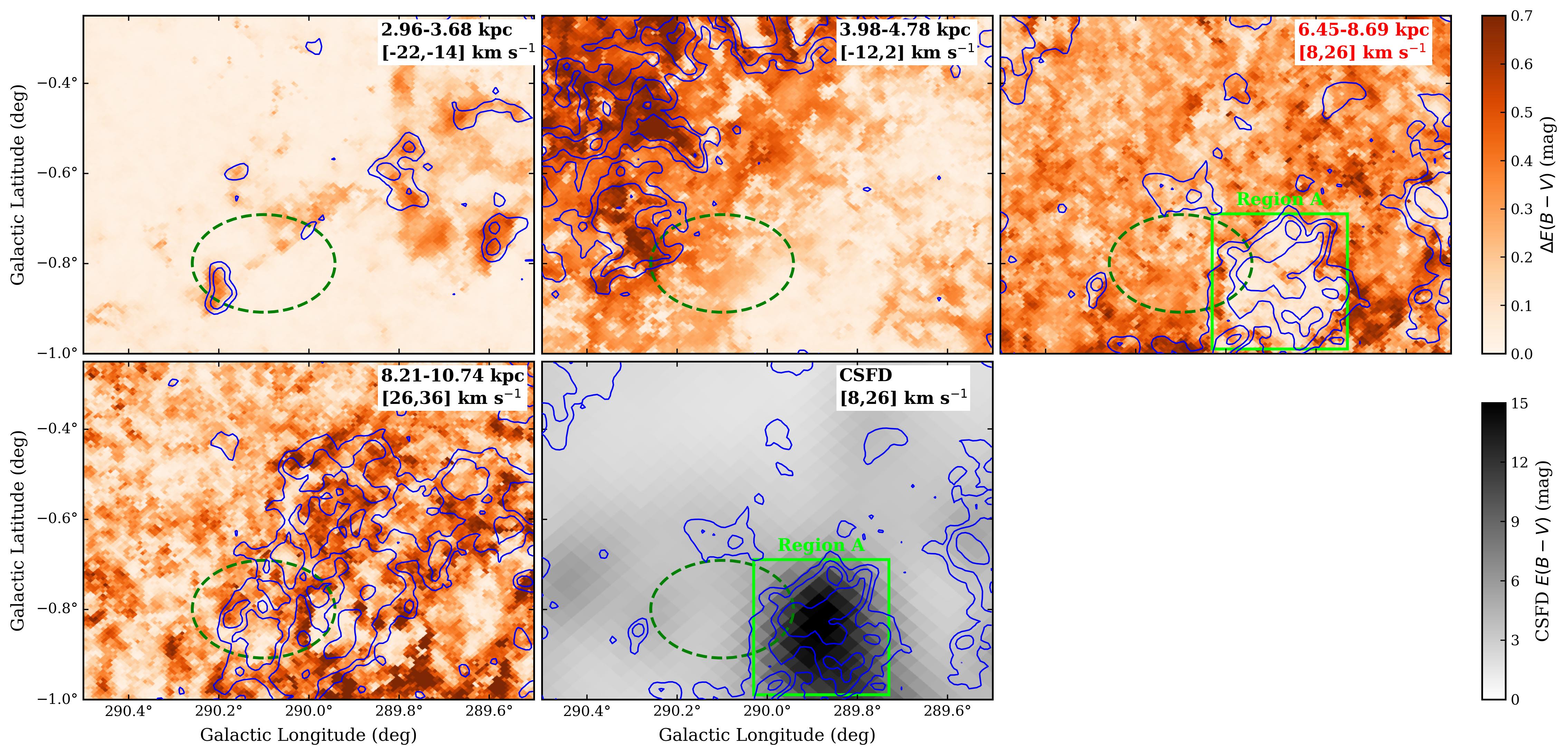}  
    \caption{Differential extinction maps ($\Delta E(B-V)$) toward G290.1$-$0.8 for distance bins corresponding to the 10\% intensity threshold of the extinction gradient peaks, along with the 2D extinction map from \protect\hyperlink{CSFD_def}{CSFD} for comparison. Blue contours overlaid on the respective panels represent the CO emission from the velocity components associated with each distance interval. The dark green ellipse denotes the approximate boundary of the SNR, while the green box (Region A) highlights the identified extinction-missing structure, likely due to high-density clouds. The red-labeled velocity and distance ranges indicate the molecular components and their corresponding distance estimates identified as being physically associated with the SNR.}  
    \label{fig 4}  
\end{figure}

To establish the distance intervals for the identified MC components, we employed a multi-step refinement process that correlates CO velocity intervals with 3D extinction structures. Initially, systematic uniform slicing of the 3D extinction map with a constant increment of 1~kpc was performed for preliminary morphological localization (see Figure~\ref{fig 2}). By evaluating the spatial correspondence between these coarse extinction maps and the CO integrated intensity distributions, we identified the initial distance anchors for each velocity component. Subsequently, representative sightlines were selected from regions with prominent extinction features to construct extinction-distance profiles and their corresponding gradient curves (see Figure~\ref{fig 3}). These constraints were further refined by defining the distance boundaries as the range where the differential extinction gradient remains above 10\% of its peak intensity. This threshold effectively characterizes both the physical thickness of the MCs and the systematic distance uncertainties inherent in the 3D extinction maps. Following this procedure, we identified four optimized distance intervals (2.96--3.68~kpc, 3.98--4.78~kpc, 6.45--8.69~kpc, and 8.21--10.74~kpc) corresponding to the primary CO velocity components (see Figure~\ref{fig 4}). 

Specifically, for the $[8, 26]$~km~s$^{-1}$ MC candidate most likely associated with the SNR, no direct high-extinction structures spatially coincident with the cloud were detected in the 6.45--8.69~kpc slice. However, a prominent dust hole (Region~A) located on the western side of the 
SNR is observed within this interval, as indicated by the green box in Figure~\ref{fig 4}. The boundary of this hole exhibits a high degree of morphological correspondence with the extension of the MC, and the high-extinction structures toward the southeast are also consistent with the MC morphology. Combined with the 2D \hyperlink{CSFD_def}{CSFD} map and empirical extinction estimations (where the peak CO integrated intensity of $\sim$50~K~km~s$^{-1}$ implies an expected extinction of $E(B-V) \approx 3.5$~mag), we confirm that Region~A is indeed a region of high extinction. The absence of an extinction signal in the 3D slices is attributed to an observational bias: behind such high extinction at this large distance, the scarcity of detected background stars makes it difficult to extract a reliable signal. This characteristic further supports the physical association between the $[8, 26]$~km~s$^{-1}$ MC and this distance interval.

The remaining identified components are categorized as either foreground or background MCs based on their morphological alignment with the differential extinction maps. In the 2.96--3.68~kpc and 3.98--4.78~kpc slices, the $[-22, -14]$~km~s$^{-1}$ and $[-12, 2]$~km~s$^{-1}$ foreground MCs exhibit excellent spatial correspondence with discrete high-extinction patches and northeastern extinction features, respectively. For the 3.98--4.78~kpc cloud, we noted a temporary extinction deficit at the peak CO integrated intensity; however, this structure reappears as a high-extinction feature in subsequent slices (e.g., 5.0--7.0~kpc). This phenomenon likely results from the extreme density of the MC core leading to a localized deficit of background stars, a bias that is effectively compensated for as the sampling volume increases. Finally, in the 8.21--10.74~kpc slice, while the residual extinction-missing structure of the $[8, 26]$~km~s$^{-1}$ MC remains visible within Region~A, a new high-extinction feature emerges in the northern region. This new feature correlates well with the $[26, 36]$~km~s$^{-1}$ MC, thereby placing this background component at a significantly greater distance than the SNR-associated $[8, 26]$~km~s$^{-1}$ candidate.

Based on the FWHM of the extinction gradient peaks, we determined the precise distances for all identified MC components, as summarized in Table \ref{tab2}. The distances for the foreground clouds are $3.36^{+0.20}_{-0.20}$~kpc (corresponding to the $[-22, -14]$~km~s$^{-1}$ MC) and $4.33^{+0.24}_{-0.15}$~kpc (corresponding to the $[-12, 2]$~km~s$^{-1}$ MC). The SNR-associated candidate, the $[8, 26]$~km~s$^{-1}$ MC, is anchored at an accurate distance of $7.32^{+0.60}_{-0.47}$~kpc. Finally, the background $[26, 36]$~km~s$^{-1}$ MC is located at a greater distance of $9.33^{+0.55}_{-0.43}$~kpc.

In summary, assuming that the MC within the velocity range of $[8, 26]$~km~s$^{-1}$ physically interacts with SNR G290.1$-$0.8, we determine the distance to the remnant to be $7.32^{+0.60}_{-0.47}$~kpc. This result is in high agreement with the kinematic distances of 7.0~kpc, 7.0--8.0~kpc, and $7.0 \pm 1.0$~kpc derived by \citet{1996A&A...315..243R}, \citet{2005SerAJ.170...47F}, and \citet{2006MNRAS.369..416R}, respectively. 

In contrast, the extinction-based distance of 3.377~kpc reported by \citet{2025ApJ...988..176C} exhibits a significant discrepancy with our findings. Through a detailed comparison of the differential extinction maps, we suggest that the bias in \citet{2025ApJ...988..176C} primarily stems from the limitations of their star sample at large distances. Due to the lack of reliable stellar data, their model failed to trace distant dust clouds beyond 5~kpc. This limitation led to a misidentification of the extinction jump caused by the foreground MC at $[-22, -14]$~km~s$^{-1}$ (located at 2.96--3.68~kpc) as the signal from the SNR itself. This finding underscores how the completeness of distant stellar samples can affect extinction-based distance estimates, while demonstrating the capability of our joint method in disentangling complex line-of-sight structures. A re-examination of early studies reveals that although \citet{1973ApL....15...61D} placed the SNR at a closer distance of 3.4--4.0~kpc based on HI absorption, that work was constrained by the uncertainties in contemporary Galactic rotation models and limited angular resolution, making it susceptible to interference from overlapping components along the line of sight. Notably, \citet{1973ApL....15...61D} utilized the H109$\alpha$ recombination line to measure the distance of the adjacent H~II region MSH11-61B as approximately 7.5~kpc, which is remarkably consistent with our result of 7.32~kpc.

Furthermore, two young pulsars in the vicinity provide independent support for a distance of $\sim$7~kpc. The dispersion measure of PSR J1105$-$6107 suggests a distance of 7~kpc according to \citet{1993ApJ...411..674T}, or 5~kpc as derived by \citet{2014ApJ...795L..27H}; however, its association with the SNR remains a subject of debate due to its large characteristic age and the absence of a clear morphological link \citep{1998ApJ...497L..29G}. In contrast, PSR J1101$-$6101 exhibits a remarkably prominent cometary X-ray tail (the Lighthouse Nebula), with a geometric alignment that points precisely toward the center of SNR G290.1$-$0.8. This configuration strongly implies that the pulsar is escaping from the remnant's center at a high velocity \citep{2014A&A...562A.122P, 2014ApJ...795L..27H, 2023ApJ...950..177K}. While studies of PSR J1101$-$6101 typically adopt the 7~kpc distance from \citet{2006MNRAS.369..416R}, it is generally held that under this distance assumption, the observed ultra-long jet and the inferred high transverse velocity are physically self-consistent within established dynamical models.

Following the multi-step procedure described above, we determined the optimized distance intervals for the various MC components across the fields of all sample SNRs by correlating their CO velocity components with the 3D extinction distribution. To avoid redundancy, this same systematic workflow is uniformly applied to the remaining nine~SNRs, with their respective results detailed in the subsequent subsections.

\subsection{G292.2$-$0.5}

\cite{2004MNRAS.352.1405C} derived a radial velocity of \(22 \pm 6\) km s\(^{-1}\) for the SNR G292.2$-$0.5 from HI absorption spectra. \cite{2019PASA...36...14V} combined CO and X-ray observations and suggested that the remnant interacts with a MC in the velocity range \([20, 30]\) km s\(^{-1}\). Based on Mopra data, we have identified four main MC components in the target region (see Figures \ref{fig 1}(b) and \ref{fig:A1}), with velocity ranges of \([-30, -16]\) km s\(^{-1}\), \([-16, 6]\) km s\(^{-1}\), \([6, 20]\) km s\(^{-1}\), and \([20, 36]\) km s\(^{-1}\). Following the studies of \cite{2004MNRAS.352.1405C} and \cite{2019PASA...36...14V}, we consider that the \([20, 36]\) km s\(^{-1}\) MC component identified here is likely physically interacting with the SNR.

\begin{figure}[htb!]
    \centering
    \includegraphics[width=0.9\textwidth]{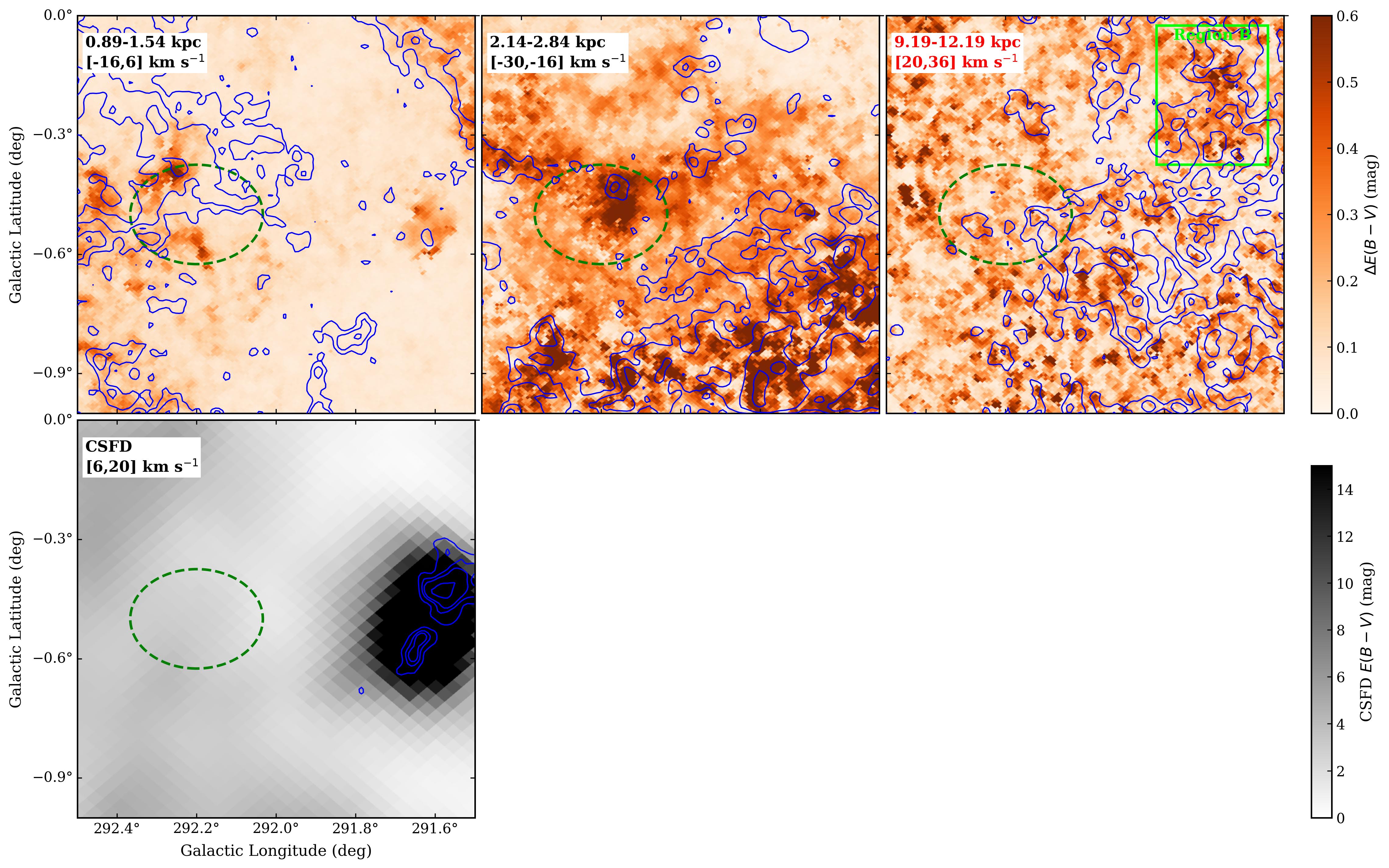}  
    \caption{Same as Figure \ref{fig 4}, but for G292.2$-$0.5.}
    \label{fig 5}  
\end{figure}

Regarding the candidate associated with the SNR, the $[20, 36]$~km~s$^{-1}$ MC is anchored within the 9.19--12.19~kpc interval (see Figure~\ref{fig 5}). Although the scarcity of background stars at such large distances limits the structural completeness of the 3D model, we identified a prominent extinction feature in Region~B (located to the northwest of the SNR, Figure~\ref{fig 5}) that spatially correlates with the cloud morphology. 

The remaining components are categorized as either foreground or background MCs. In the 0.89--1.54~kpc interval, the extinction structures in the northwest and southeast corners exhibit good morphological agreement with the $[-16, 6]$~km~s$^{-1}$ MC, though some discrepancies remain between the eastern structure of the MC and the observed extinction. In the 2.14--2.84~kpc interval, the high-extinction structures are highly consistent with the intense CO emission of the $[-30, -16]$~km~s$^{-1}$ MC. Finally, for the $[6, 20]$~km~s$^{-1}$ MC, the peak CO integrated intensity reaches $\sim$120~K~km~s$^{-1}$ (implying $E(B-V) \approx 8.4$~mag). Despite its prominence in the 2D \hyperlink{CSFD_def}{CSFD} map, no matching high-extinction structures or extinction-missing phenomena were detected in any differential slices within 13~kpc. This confirms that this component is a distant background cloud located beyond the 13~kpc detection limit of the current 3D dust map.

Based on the FWHM of the extinction gradient peaks, we determined the precise distances for all identified components, as summarized in Table~\ref{tab2}. \textbf{The foreground clouds are placed at $1.35^{+0.12}_{-0.17}$~kpc (corresponding to the $[-16, 6]$~km~s$^{-1}$ MC) and $2.45^{+0.19}_{-0.20}$~kpc (corresponding to the $[-30, -16]$~km~s$^{-1}$ MC), respectively.} The SNR-associated $[20, 36]$~km~s$^{-1}$ MC is anchored at a significantly greater distance of $10.85^{+0.43}_{-0.68}$~kpc.

Consequently, we determine the distance to SNR G292.2$-$0.5 to be $10.85^{+0.43}_{-0.68}$~kpc, which corresponds to the distance of the MC in the velocity range of $[20, 36]$~km~s$^{-1}$ that potentially interacts with the remnant. Our result is broadly consistent with the kinematic distances of $8.4 \pm 0.4$~kpc obtained by \citet{2004MNRAS.352.1405C} and 8.6--9.7~kpc derived by \citet{2019PASA...36...14V}. In contrast, the distances estimated by \citet[3.6--6.3~kpc]{2005ApJ...619..856G} and \citet[2.422~kpc]{2025ApJ...988..176C} show significant discrepancies with our value. We attribute these differences primarily to the complex extinction structures along this line of sight. Specifically, we suggest that the extinction distance reported by \citet{2025ApJ...988..176C} should be assigned to the foreground MC within the velocity range of $[-30, -16]$~km~s$^{-1}$ located at 2.14--2.84~kpc.

\subsection{G296.1$-$0.5}

G296.1$-$0.5 is a complex shell-type SNR whose southwestern side exhibits a double-shell structure in both radio and X-ray bands. \cite{1977MNRAS.181..541L} inferred a LSR velocity of $-35$~km~s\(^{-1}\) for the remnant based on H$\alpha$ and H$\beta$ line observations. \cite{2022ApJ...933..101T} combined HI, CO, and X-ray observations and, using morphological and kinematic features, identified CO and HI gas in the velocity range \([-30, -15]\)~km~s\(^{-1}\) as associated with the remnant. Based on Mopra data, we have identified five main MC components with velocity intervals of \([-40, -34]\)~km~s\(^{-1}\), \([-34, -24]\)~km~s\(^{-1}\), \([-18, -14]\)~km~s\(^{-1}\), \([-8, 0]\)~km~s\(^{-1}\), and \([10, 28]\)~km~s\(^{-1}\) (see Figures \ref{fig 1}(c) and \ref{fig:A1}). Among these, the MC in the velocity range \([-34, -24]\)~km~s\(^{-1}\) exhibits a clear arc-like structure and is spatially distinct from the \([-18, -14]\)~km~s\(^{-1}\) cloud. Following the analysis of \cite{2022ApJ...933..101T}, we consider that the \([-34, -24]\)~km~s\(^{-1}\) MC is physically interacting with the SNR.

\begin{figure}[htb!]
    \centering
    \includegraphics[width=0.9\textwidth]{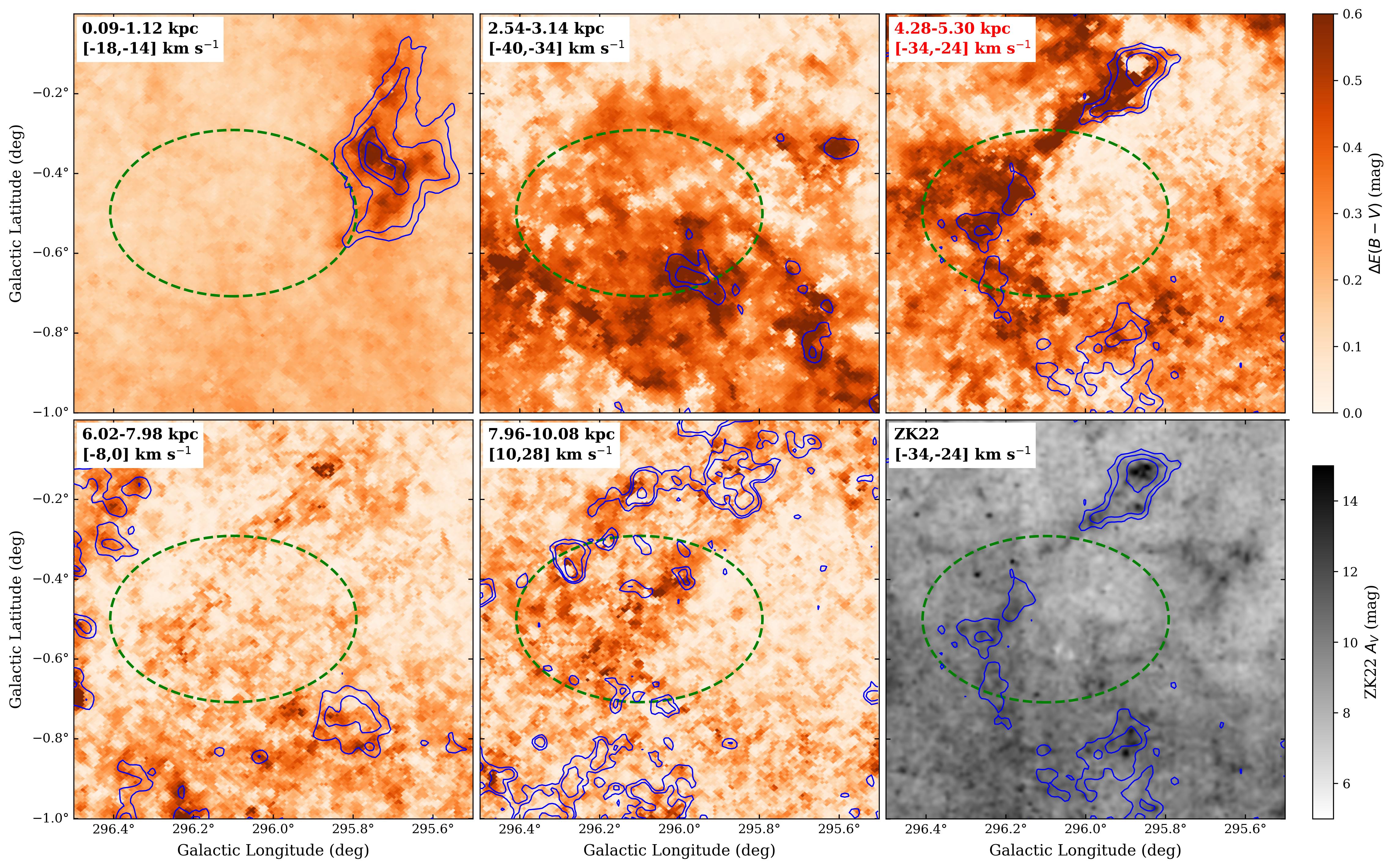}  
    \caption{Same as Figure \ref{fig 4}, but for G296.1$-$0.5. Note that the 2D extinction map in the final panel is from \protect\hyperlink{ZK22_definition}{ZK22}.}
    \label{fig 6}  
\end{figure}

Regarding the candidate associated with the SNR, the $[-34, -24]$~km~s$^{-1}$ MC is anchored within the 4.28--5.30~kpc interval (see Figure~\ref{fig 6}). In the corresponding differential slice, an arc-like high-extinction structure extending from the north to the east shows an excellent morphological match with the cloud. Notably, an extinction-missing feature is observed at the peak of the northern CO emission (where the integrated intensity of $\sim$20~K~km~s$^{-1}$ implies an expected $E(B-V) \approx 1.4$~mag). Despite this gap in the 3D model, this same location exhibits intense extinction in the 2D \hyperlink{ZK22_definition}{ZK22} extinction map. This confirms that the missing feature is an observational limitation caused by extreme extinction at this distance, which leads to a deficit of background stars and prevents the extraction of a reliable 3D signal. This characteristic further reinforces the physical association between the $[-34, -24]$~km~s$^{-1}$ MC and this distance interval.

The remaining components are categorized as either foreground or background MCs. In the 0.09--1.12~kpc interval, the western extinction structures exhibit a high degree of morphological correspondence with the $[-18, -14]$~km~s$^{-1}$ MC. In the 2.14--2.84~kpc interval, three prominent high-extinction centers accurately trace the features of the $[-40, -34]$~km~s$^{-1}$ MC, although the overall extinction distribution is broader than the CO emission. Further out, the 6.02--7.98~kpc interval shows extinction structures in the east and south consistent with the $[-8, 0]$~km~s$^{-1}$ MC. Finally, in the 7.96--10.08~kpc interval, the northern arc and southern structures correlate well with the $[10, 28]$~km~s$^{-1}$ MC.

Based on the FWHM of the extinction gradient peaks, we determined the precise distances for all identified components, as summarized in Table~\ref{tab2}. The identified MCs are located at the following distances: $0.80^{+0.14}_{-0.16}$~kpc (corresponding to the $[-18, -14]$~km~s$^{-1}$ MC), $2.78^{+0.15}_{-0.11}$~kpc (corresponding to the $[-40, -34]$~km~s$^{-1}$ MC), $6.83^{+0.68}_{-0.52}$~kpc (corresponding to the $[-8, 0]$~km~s$^{-1}$ MC), and $8.96^{+0.51}_{-0.49}$~kpc (corresponding to the $[10, 28]$~km~s$^{-1}$ MC). The SNR-associated candidate, the $[-34, -24]$~km~s$^{-1}$ MC, is anchored at an accurate distance of $4.59^{+0.18}_{-0.19}$~kpc.

Consequently, assuming a physical association between the MC in the velocity range of $[-34, -24]$~km~s$^{-1}$ and SNR G296.1$-$0.5, the distance to the remnant is determined to be $4.59^{+0.18}_{-0.19}$~kpc. This result is in good agreement with the distance estimate of 4.9~kpc derived from the $\Sigma$--$D$ relationship by \citet{1973Natur.246...28C}. Our findings are also broadly consistent with the upper limits provided in previous studies, including the limits of $3 \pm 1$~kpc and 4~kpc obtained from reddening measurements and kinematics, respectively, by \citet{1977MNRAS.181..541L}, as well as the upper bound of $3.8 \pm 0.5$~\text{kpc} derived from red clump star extinction by \citet{2020A&A...639A..72W}. However, our result exhibits a significant discrepancy with the kinematic distance of $2.1^{+0.9}_{-0.6}$~kpc inferred by \citet{2022ApJ...933..101T}. We suggest that this deviation may be attributed to prominent non-circular motions along this line of sight, which could substantially compromise the reliability of traditional kinematic distance determinations.

\subsection{G296.8$-$0.3}

G296.8$-$0.3 is a complex SNR with multiple shell structures. While direct physical evidence of a localized interaction with molecular gas (e.g., CO line broadening or OH masers) has not been definitively reported, the remnant possesses a reliable kinematic constraint from HI observations. Based on HI absorption spectra, \citet{1998MNRAS.296..813G} determined the systemic velocity of the remnant to be within $[15, 30]$~km~s$^{-1}$. Following the selection criteria in Section 2.3, this kinematic information provides the physical basis for associating the remnant with its surrounding ISM. Using Mopra data, we identified three main MC components in the region: $[-38, -24]$~km~s$^{-1}$, $[-12, 2]$~km~s$^{-1}$, and $[8, 36]$~km~s$^{-1}$ (see Figures~\ref{fig 1}d and \ref{fig:A1}). The first two MC components are spatially distant from the SNR. In contrast, the $[8, 36]$~km~s$^{-1}$ component is not only spatially close to the remnant's shell but also exhibits significant velocity overlap with the HI-derived systemic range of $[15, 30]$~km~s$^{-1}$ \citep{1998MNRAS.296..813G}. Therefore, we adopt this MC as the target component for our distance analysis.

\begin{figure}[htb!]
    \centering
    \includegraphics[width=0.9\textwidth]{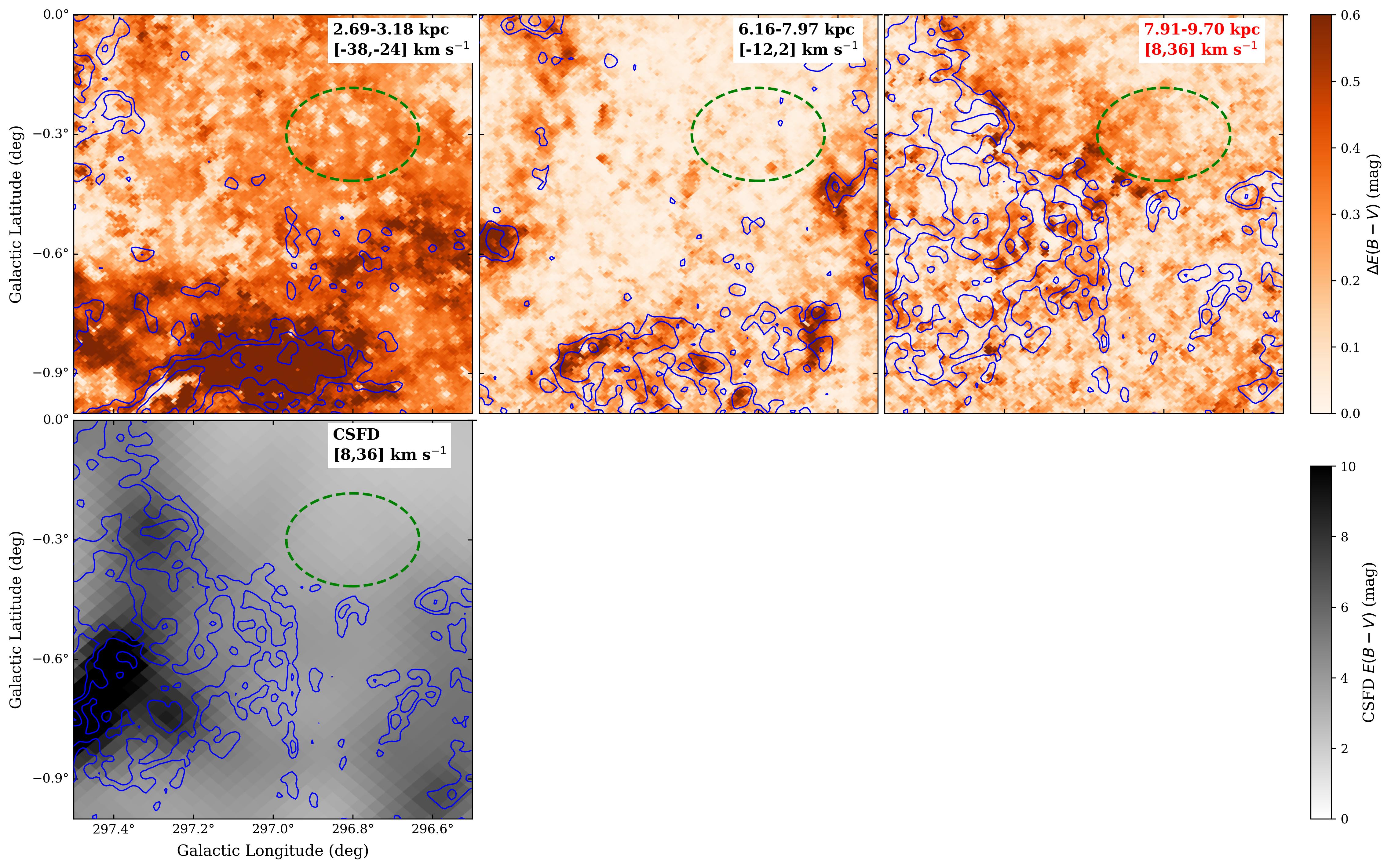}  
    \caption{Same as Figure \ref{fig 4}, but for G296.8$-$0.3.}

    \label{fig 7}  
\end{figure}

Regarding the candidate associated with the SNR, the $[8, 36]$~km~s$^{-1}$ MC is anchored within the 7.91--9.70~kpc interval (see Figure~\ref{fig 7}). In the corresponding differential slice, while the overall extinction structure exhibits significant complexity, the high-extinction features in the central part of the slice correlate well with the peripheral regions of the MC, where the relatively weak CO emission ($\sim$10~K~km~s$^{-1}$) corresponds to $E(B-V) \approx 0.7$~mag. Yet, an extensive extinction-missing structure is observed toward the eastern main body of the MC. In this region, the intense CO emission (20--50~K~km~s$^{-1}$) implies a high expected extinction of $E(B-V) \approx 1.4$--$3.5$~mag. Comparisons with the \hyperlink{CSFD_def}{CSFD} 2D extinction map confirm the presence of high extinction in this eastern part. This confirms that the missing signal is an observational bias stemming from the scarcity of detected background stars at such a large distance, which are insufficient to resolve the high extinction of the cloud's main body. Consequently, this characteristic supports placing the $[8, 36]$~km~s$^{-1}$ MC within this distance interval.

The remaining components are categorized as foreground MCs. In the 2.69--3.18~kpc interval, the southern high-extinction region shows good morphological agreement with the $[-38, -24]$~km~s$^{-1}$ MC. In the 6.16--7.97~kpc interval, all identified high-extinction structures are highly consistent with the morphology of the $[-12, 2]$~km~s$^{-1}$ MC.

Based on the FWHM of the extinction gradient peaks, we determined the precise distances for all identified components, as summarized in Table~\ref{tab2}. The foreground clouds are placed at $2.91^{+0.15}_{-0.11}$~kpc (associated with the $[-38, -24]$~km~s$^{-1}$ MC) and $7.12^{+0.50}_{-0.62}$~kpc (associated with the $[-12, 2]$~km~s$^{-1}$ MC), respectively. The SNR-associated candidate, the $[8, 36]$~km~s$^{-1}$ MC, is anchored at an accurate distance of $8.74^{+0.40}_{-0.29}$~kpc.

Accordingly, we determine the distance to SNR G296.8$-$0.3 to be $8.74^{+0.40}_{-0.29}$~kpc, based on its physical association with the $[8, 36]$~km~s$^{-1}$ MC. This result is in excellent agreement with the kinematic distance of $9.6 \pm 0.6$~kpc derived from HI absorption \citep{1998MNRAS.296..813G} and the lower limit of $>9$~kpc inferred from X-ray absorption column density \citep{2012Ap&SS.337..573S}. In contrast, our findings exhibit a significant discrepancy with the distance of 3.294~kpc reported by \citet{2025ApJ...988..176C}. We suggest that this nearby signal primarily stems from the foreground MC in the velocity range of $[-38, -24]$~km~s$^{-1}$ (located at 2.69--3.18~kpc). In this interpretation, the extinction jump identified in \citet{2025ApJ...988..176C} reflects the presence of this foreground component rather than the more distant SNR. 

\subsection{G298.6$-$0.0}

G298.6$-$0.0 exhibits a well-defined radio shell structure. \citet{2016PASJ...68S...5B} detected centrally filled X-ray emission within its radio shell and thus classified it as a mixed-morphology SNR. Based on CO and HI data, \citet{2023PASJ...75..384Y} suggested that the MC in the velocity range \([19, 34]\)~km~s\(^{-1}\) may be physically associated with the remnant. With Mopra observations, we identified five main MC components in the target region, with velocity intervals of \([-42, -28]\)~km~s\(^{-1}\), \([-28, -24]\)~km~s\(^{-1}\), \([-22, -8]\)~km~s\(^{-1}\), \([4, 10]\)~km~s\(^{-1}\), and \([14, 36]\)~km~s\(^{-1}\) (see Figures~\ref{fig 1}(e) and \ref{fig:A1}). Following the analysis of \citet{2023PASJ...75..384Y}, we consider the \([14, 36]\)~km~s\(^{-1}\) MC to be likely physically interacting with the remnant, and adopt it as the target cloud for subsequent analysis.

\begin{figure}[htb!]
    \centering
    \includegraphics[width=0.9\textwidth]{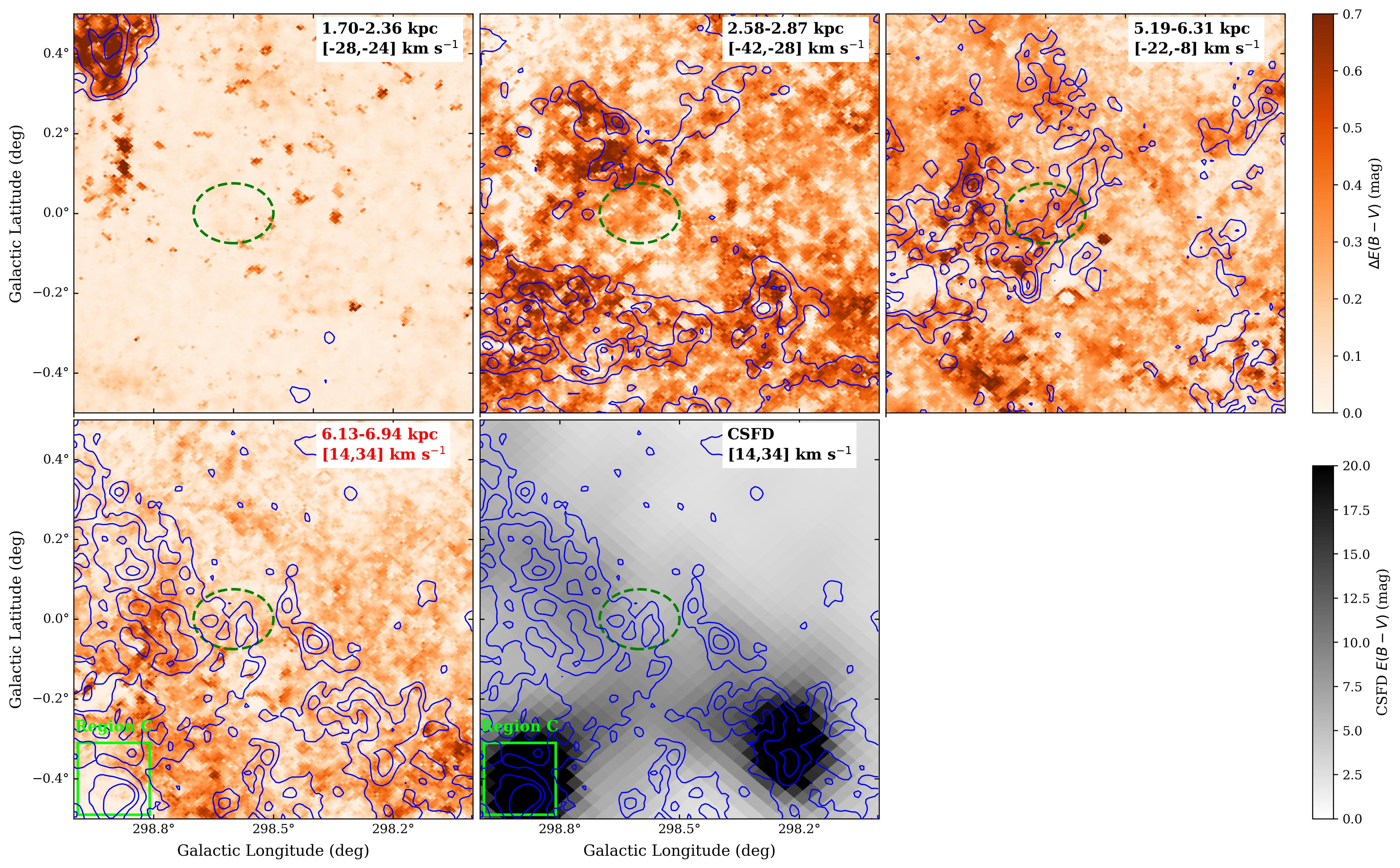}  
    \caption{Same as Figure \ref{fig 4}, but for G298.6$-$0.0.}
    \label{fig 8}  
\end{figure}

Regarding the candidate associated with the SNR, the $[14, 34]$~km~s$^{-1}$ MC is anchored within the 6.13--6.94~kpc interval (see Figure~\ref{fig 8}). In the corresponding differential slice, the southwestern extinction structure exhibits good agreement with the cloud morphology. Although some morphological discrepancies exist in the eastern part, we identified a significant extinction-missing feature in Region~C (southeast corner, Figure~\ref{fig 8}), which is spatially coincident with the cloud's core. The peak integrated CO intensity reaches $\sim$150~K~km~s$^{-1}$, implying an extreme expected extinction of $E(B-V) \approx 10.5$~mag. Verified by the \hyperlink{CSFD_def}{CSFD} 2D extinction map, this feature is confirmed to arise from extreme extinction. This observational bias occurs because the high-extinction core lacks enough background stars to generate a reliable 3D signal at such a large distance. When combined with the spatial coupling observed in the southwestern structure, this characteristic justifies placing the $[14, 34]$~km~s$^{-1}$ MC within the 6.13--6.94~kpc interval.

The remaining components are categorized as foreground or background MCs. In the 1.70--2.36~kpc interval, a high-extinction structure in the northeast perfectly matches the $[-28, -24]$~km~s$^{-1}$ MC. In the 2.58--2.87~kpc interval, the extinction structures in the south and north-central parts correlate well with the $[-42, -28]$~km~s$^{-1}$ MC. Further out, in the 5.19--6.31~kpc interval, the eastern extinction features match the $[-22, -8]$~km~s$^{-1}$ MC, with an extinction-missing structure corresponding to its intense CO emission. Regarding the $[4, 10]$~km~s$^{-1}$ MC, no associated extinction structures were detected within 10~kpc. Its spatial distribution overlaps significantly with the dense regions of the target $[14, 34]$~km~s$^{-1}$ MC; this spatial coincidence likely leads to a blending of extinction features, which potentially complicates or hinders an independent distance determination for this component.

Based on the FWHM of the extinction gradient peaks, we determined the precise distances for all identified components, as summarized in Table~\ref{tab2}. The foreground clouds are placed at $1.94^{+0.26}_{-0.14}$~kpc (corresponding to the $[-28, -24]$~km~s$^{-1}$ MC), $2.76^{+0.07}_{-0.09}$~kpc (corresponding to the $[-42, -28]$~km~s$^{-1}$ MC), and $5.83^{+0.24}_{-0.34}$~kpc (corresponding to the $[-22, -8]$~km~s$^{-1}$ MC), respectively. The SNR-associated $[14, 34]$~km~s$^{-1}$ MC is anchored at an accurate distance of $6.50 \pm 0.21$~kpc.

Therefore, we assign the SNR G298.6$-$0.0, which is potentially associated with the MC in the velocity range \([14, 34]\)~km~s\(^{-1}\), a distance of \(6.50^{+0.21}_{-0.21}\)~kpc. This result is closer than the kinematic distance of \(10.1 \pm 0.5\)~kpc obtained by \citet{2023PASJ...75..384Y}. We suggest that this discrepancy likely reflects the presence of local non-circular motions in this region, which can affect kinematic distance estimates derived from rotation curve models.

\subsection{G312.4$-$0.4}
G312.4$-$0.4 is a shell-type SNR displaying a characteristic horseshoe-shaped radio morphology. It was first discovered in the MOST 408 MHz survey by \citet{1985MNRAS.216..753C}. Based on the H$_2$CO molecular absorption features, they identified a H$_2$CO cloud with a velocity of about $-49$~km~s$^{-1}$ near the remnant. \citet{1996AJ....111.1651F} detected 1720 MHz hydroxyl (OH) maser emission with the Parkes 64\,m telescope, suggesting a possible interaction between the remnant and a MC. \citet{2003MNRAS.339.1048D} analyzed HI absorption spectra and proposed a systemic velocity range of $[20, 40]$~km~s$^{-1}$ for the remnant. However, \citet{2022ApJ...940...63R} re-examined the HI absorption data and found a tangential velocity along the line of sight of about $-65$~km~s$^{-1}$, and noted that the absorption feature does not extend to the tangent point. They therefore argued that the systemic velocity should be around $-50$~km~s$^{-1}$. \citet{2023ApJ...959...97C}, referencing the Milky Way MC catalog of \citet{2016ApJ...822...52R}, identified a dense MC at 3.68~kpc whose spatial position shows a strong overlap with the $\gamma$-ray emission region of the remnant, further supporting a physical interaction between the SNR and MC.

\begin{figure}[htb!]
    \centering
    \includegraphics[width=0.9\textwidth]{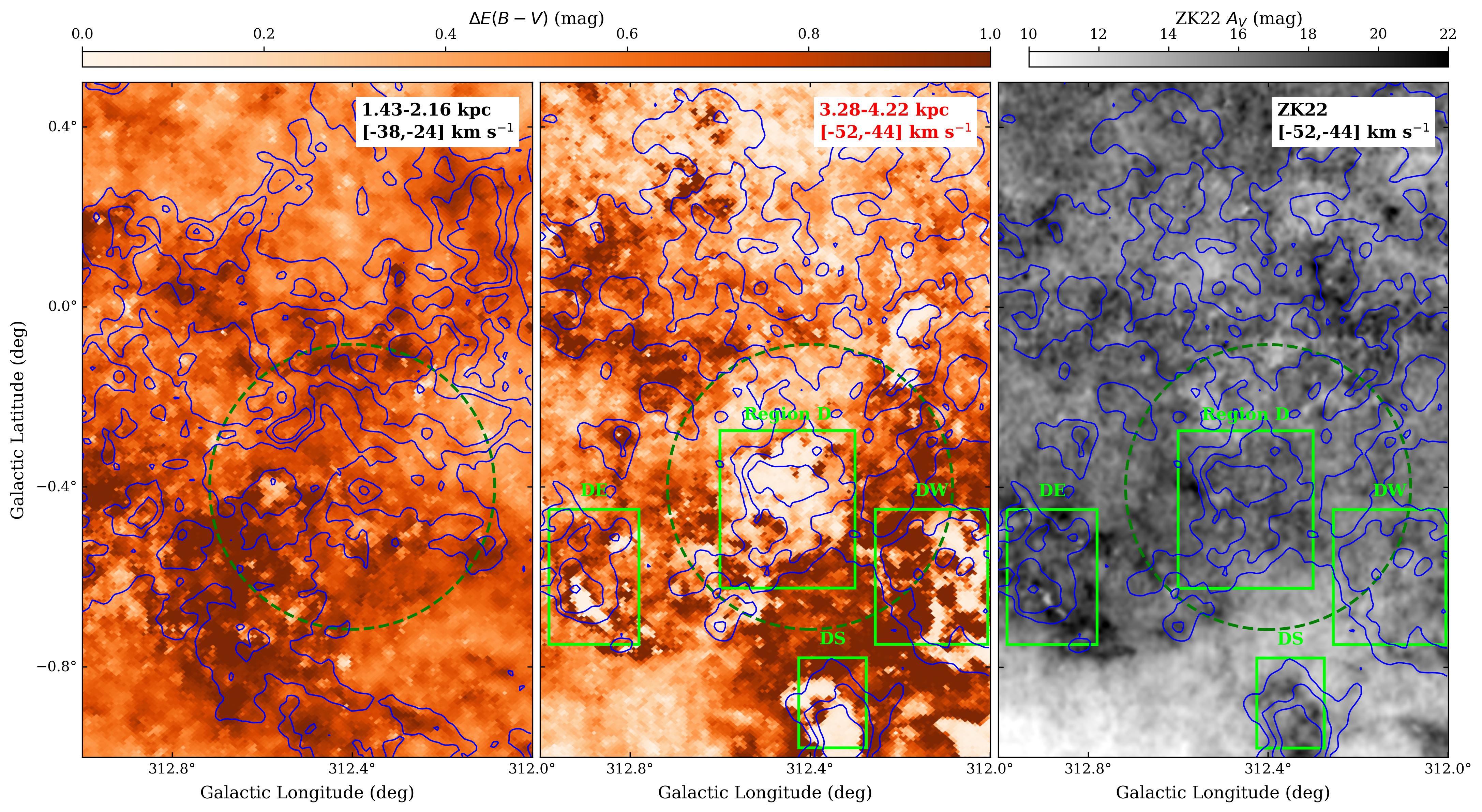}  
    \caption{Same as Figure \ref{fig 4}, but for G312.4$-$0.4. Note that the 2D extinction map in the final panel is from \protect\hyperlink{ZK22_definition}{ZK22}.}
    \label{fig 9}  
\end{figure}

Using Mopra observations, we have identified four main MC components in the region (see Figures~\ref{fig 1}(f) and \ref{fig:A2}), with velocity ranges of $[-70, -60]$~km~s$^{-1}$, $[-52, -44]$~km~s$^{-1}$, $[-38, -24]$~km~s$^{-1}$, and $[-20, 0]$~km~s$^{-1}$. In addition, using the velocity information from \citet{2003MNRAS.339.1048D}, we also identified a component in the velocity interval of $[20, 40]$~km~s$^{-1}$, whose CO emission is relatively weak and spatially diffuse. The MC in the velocity range $[-52, -44]$~km~s$^{-1}$, in particular its substructure Region~D (Figure~\ref{fig 9}), is spatially nestled within the depression of the remnant’s horseshoe-shaped radio emission (see Figure \ref{fig:A2}). This pronounced morphological coincidence, together with the results of \citet{2022ApJ...940...63R} and \citet{2023ApJ...959...97C}, leads us to conclude that this MC is likely physically interacting with the SNR.

Regarding the candidate associated with the SNR, the $[-52, -44]$~km~s$^{-1}$ MC is anchored within the 3.28--4.22~kpc interval (see Figure~\ref{fig 9}). In the corresponding differential slice, significant extinction-missing features are observed in four sub-regions: D, DE, DS, and DW. These regions spatially coincide with the peaks of CO integrated intensity ($\sim$55~K~km~s$^{-1}$), which imply a high expected extinction of $E(B-V) \approx 3.85$~mag. Comparisons with the \hyperlink{ZK22_definition}{ZK22} 2D extinction map confirm that the 3D signal deficit in these four regions originates from extreme extinction that hinders the detection of sufficient background stars at this distance. Furthermore, in the eastern and southwestern parts of the slice, matching high-extinction structures can be traced in regions with relatively weaker CO emission. This dual evidence—the morphological coupling in lower-extinction regions and the consistent observational bias in high-extinction cores—firmly places the $[-52, -44]$~km~s$^{-1}$ MC within the 3.28--4.22~kpc interval.

The remaining components are categorized as foreground or background MCs.In the 1.43--2.16~kpc interval, the extinction structures exhibit a broad distribution, with the southern and northern high-extinction features agreeing morphologically with the $[-38, -24]$~km~s$^{-1}$ MC. Regarding the $[-70, -60]$~km~s$^{-1}$, $[-20, 0]$~km~s$^{-1}$, and $[20, 40]$~km~s$^{-1}$ MC components, no associated extinction structures or significant extinction jumps were detected within 10~kpc.

Based on the FWHM of the extinction gradient peaks, we determined the precise distances for all identified components, as summarized in Table~\ref{tab2}. The foreground $[-38, -24]$~km~s$^{-1}$ MC is placed at $1.96^{+0.10}_{-0.16}$~kpc, while the SNR-associated $[-52, -44]$~km~s$^{-1}$ MC is anchored at an accurate distance of $3.60^{+0.19}_{-0.23}$~kpc.

Consequently, we determine the distance to SNR G312.4$-$0.4 to be $3.60^{+0.19}_{-0.23}$~kpc, which corresponds to the distance of its potentially interacting MC in the velocity range of $[-52, -44]$~km~s$^{-1}$. This result is in good agreement with previous studies within the reported uncertainties, including the lower-limit distance of 3.8~kpc derived from H$_2$CO absorption lines by \citet{1985MNRAS.216..753C}, and the kinematic distance of $3.5 \pm 0.5$~kpc obtained through the re-analysis of HI absorption spectra by \citet{2022ApJ...940...63R}. Furthermore, our findings are consistent with the distance lower limit of $4.41 \pm 0.5$~kpc estimated via the red clump star extinction method by \citet{2020A&A...639A..72W}. Regarding physical modeling constraints, our measurement is consistent with the upper limit of $4.0$~kpc derived from hadronic model fitting in \citet{2023ApJ...959...97C}. Furthermore, our determined distance of $3.60^{+0.19}_{-0.23}$~kpc exhibits excellent agreement with the 3.68~kpc distance for the dense MC proposed in the same study. This alignment provides independent observational support for their physical interpretation of the SNR's environment.

\subsection{G332.4$-$0.4}

\begin{figure}[htb!]
    \centering
    \includegraphics[width=0.9\textwidth]{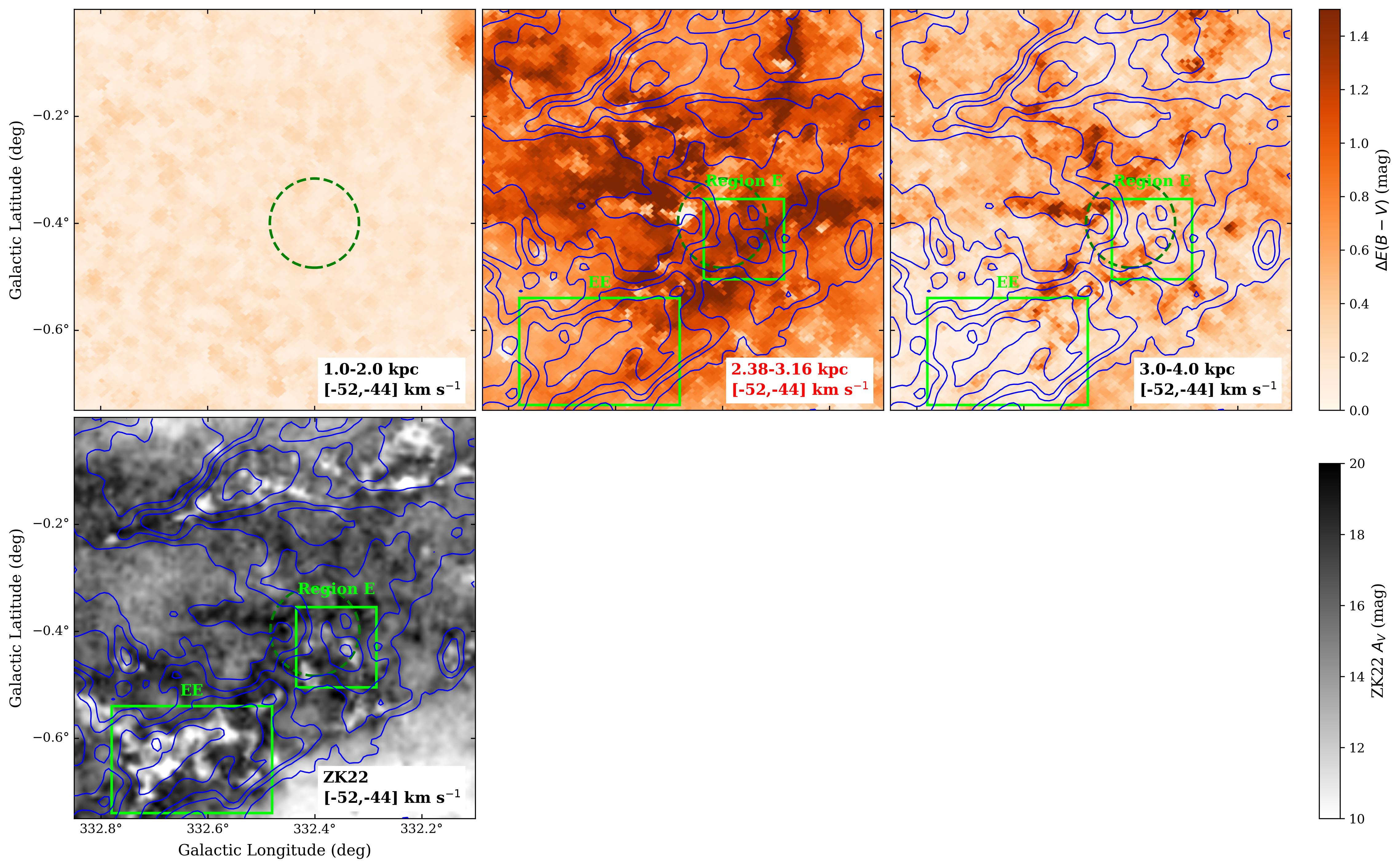}  
    \caption{Same as Figure \ref{fig 4}, but for G332.4$-$0.4. Note that the 2D extinction map in the final panel is from \protect\hyperlink{ZK22_definition}{ZK22}.}
    \label{fig 10}  
\end{figure}

G332.4$-$0.4 is a shell-type SNR that exhibits a complete and symmetric circular shell structure in the radio band. \cite{2006PASA...23...69P} constrained the systemic velocity of the remnant to $-48$~km~s$^{-1}$ through an analysis of HCO$^+$ and CO emission lines, and noted that the remnant is physically interacting with a MC to its south. Using Mopra observations, we identified three main MC components in the region, with velocity ranges of \([-110, -80]\)~km~s\(^{-1}\), \([-76, -66]\)~km~s\(^{-1}\), and \([-52, -44]\)~km~s\(^{-1}\) (see Figures~\ref{fig 1}(g) and \ref{fig:A2}). Following the study of \citet{2006PASA...23...69P}, we confirm that the Region~E (Figure~\ref{fig 10}) structure within the \([-52, -44]\)~km~s\(^{-1}\) MC (see Figure \ref{fig:A2}) is a dense molecular clump directly interacting with the SNR. Given the relatively small spatial scale of this clump, we adopt the entire MC in the velocity range \([-52, -44]\)~km~s\(^{-1}\) as the target component for our study.

Regarding the candidate associated with the SNR, the $[-52, -44]$~km~s$^{-1}$ MC is anchored within the 2.38--3.16~kpc interval (see Figure~\ref{fig 10}). In this slice, the overall extinction morphology exhibits significant discrepancies with the target MC, yet we identified a prominent extinction-missing structure in Region~EE (southeast corner). Therein, the peak integrated CO intensity reaches $\sim$90~K~km~s$^{-1}$, implying an expected extinction of $E(B-V) \approx 6.3$~mag. This level of extinction leads to a near-total signal deficit, manifesting as an extinction-missing feature not only in the 3D model but also in the \hyperlink{ZK22_definition}{ZK22} 2D map. We suggest that the extinction structures detected in this interval primarily trace the surface layer of the MC, while the dense main body remains unresolved due to the complete obscuration of background stars. To corroborate this interpretation, we examined the 3.0--4.0~kpc slice (Figure~\ref{fig 10}), which serves as a background reference. This slice exhibits a persistent and extensive extinction-missing structure, confirming that the line of sight is entirely obscured by the high-extinction dense MC located in the foreground. In contrast, the 1.0--2.0~kpc slice shows almost no extinction features, confirming that the cloud is not located at a closer distance. Consequently, based on the spatial coupling of the surface extinction and the consistent signal deficit at larger distances, we anchor the $[-52, -44]$~km~s$^{-1}$ MC within the 2.38--3.16~kpc interval. Using the FWHM of the extinction gradient peak, we determine its distance to be $2.66^{+0.23}_{-0.15}$~kpc. 

Regarding the $[-110, -80]$~km~s$^{-1}$ and $[-76, -66]$~km~s$^{-1}$ MC components, no matching extinction features or significant extinction jumps were detected within 10~kpc.

Consequently, we determine the distance to SNR G332.4$-$0.4, which is likely interacting with the target MC in the velocity range of $[-52, -44]$~km~s$^{-1}$, to be $2.66^{+0.23}_{-0.15}$~kpc. Our result is closer than the distances previously estimated by \citet{1975A&A....45..239C} and \citet{2004PASA...21...82R} based on HI absorption spectra, approximately 3.3~kpc and 3.1~kpc, respectively, as well as the kinematic distance of 3.3~kpc obtained by \citet{2006PASA...23...69P}. The disagreement likely reflects the systematic uncertainties in kinematic modeling, particularly the impact of non-circular gas dynamics which often compromises distance accuracy in this region of the Galaxy. Furthermore, our result is consistent with the distance of $3.0 \pm 0.3$~kpc derived from red clump star extinction by \citet{2019RAA....19...92S} within the uncertainties, and exhibits a good agreement with the extinction distance of 2.504~kpc reported by \citet{2025ApJ...988..176C} based on the extinction jump.

\subsection{G335.2$+$0.1}

G335.2$+$0.1 is a SNR exhibiting a filamentary shell structure. \citet{2011A&A...526A..82E} pointed out that the MC and HI gas in the velocity range \([-27, -18]\)~km~s\(^{-1}\) may be interacting with the remnant, and they observed a density depression in the HI gas coincident with the spatial location of the remnant. More recently, \citet{2025ApJ...990..213H} combined morphological, kinematic, and $\gamma$-ray features and suggested that the CO MC in the velocity range \([-48, -43]\)~km~s\(^{-1}\) may be associated with the remnant. Based on Mopra observations, we identified five main MC components in the region, with velocity intervals of \([-120, -100]\)~km~s\(^{-1}\), \([-96, -82]\)~km~s\(^{-1}\), \([-76, -56]\)~km~s\(^{-1}\), \([-52, -40]\)~km~s\(^{-1}\), and \([-28, -18]\)~km~s\(^{-1}\) (see Figures~\ref{fig 1}(h) and \ref{fig:A2}). Integrating the studies of \citet{2011A&A...526A..82E} and \citet{2025ApJ...990..213H}, we adopt both the \([-52, -40]\)~km~s\(^{-1}\) and \([-28, -18]\)~km~s\(^{-1}\) components as potential target MC that may be interacting with the remnant.

\begin{figure}[htb!]
    \centering
    \includegraphics[width=0.9\textwidth]{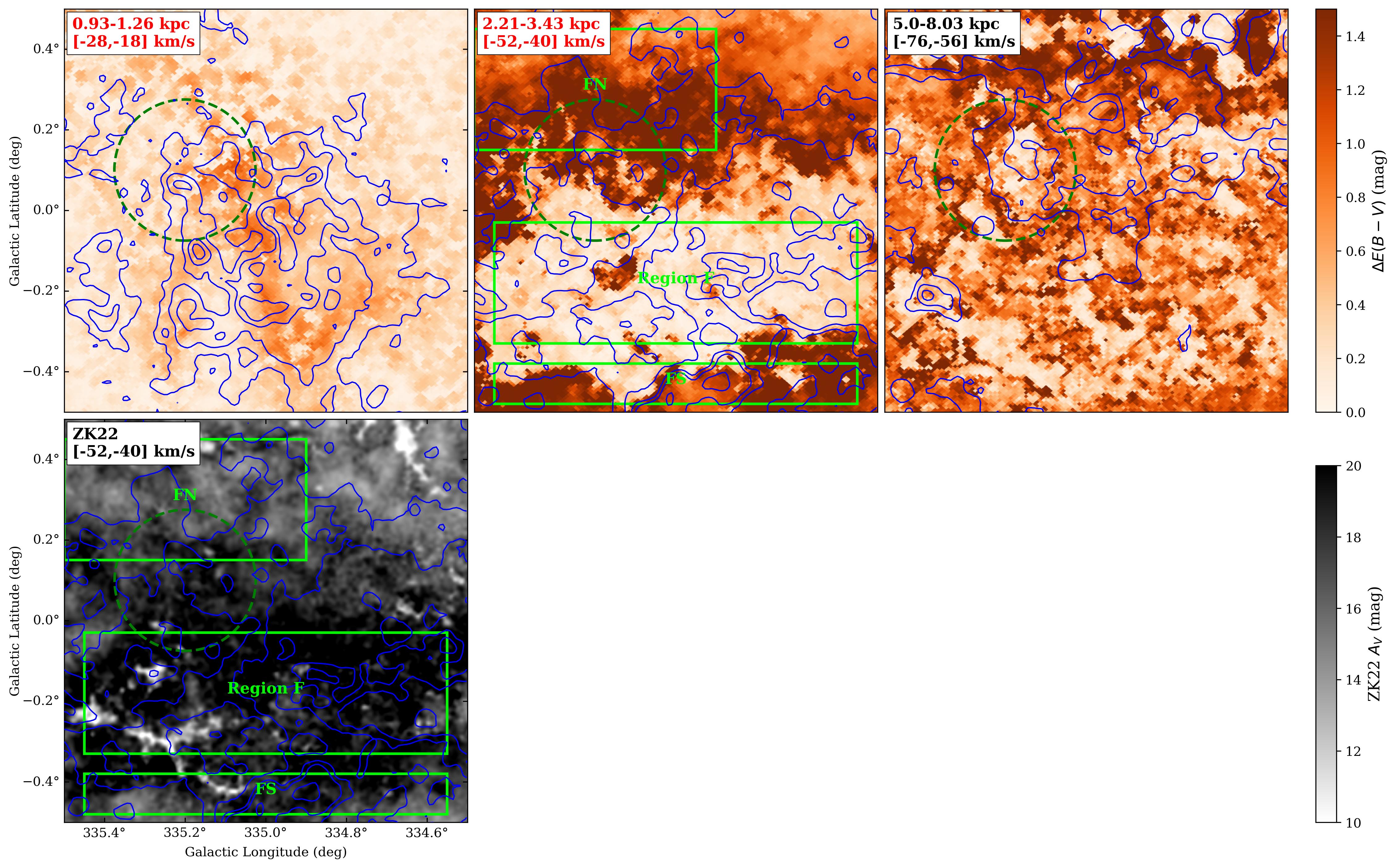}  
    \caption{Same as Figure \ref{fig 4}, but for G335.2$+$0.1. Note that the 2D extinction map in the final panel is from \protect\hyperlink{ZK22_definition}{ZK22}.}
    \label{fig 11}  
\end{figure}

 In the 0.93--1.26~kpc interval, high-extinction structures exhibit excellent morphological agreement with the $[-28, -18]$~km~s$^{-1}$ MC (see Figure~\ref{fig 11}). For the $[-52, -40]$~km~s$^{-1}$ component, the corresponding 2.21--3.43~kpc slice reveals a significant extinction-missing structure in Region~F (Figure~\ref{fig 11}), located in the south-central part of the remnant. The integrated CO intensity in this region (35--80~K~km~s$^{-1}$, implying $E(B-V) \approx 2.45$--$5.6$~mag) spatially coincides with a 3D signal deficit, independently confirmed by the extreme extinction in the \hyperlink{ZK22_definition}{ZK22} 2D map. In the regions immediately adjacent to the north and south of Region~F (Regions~FN and FS, respectively), visible extinction structures emerge where the CO emission is relatively weaker. This spatial configuration makes the entire MC appear embedded within the extinction distribution, providing a robust constraint for anchoring the $[-52, -40]$~km~s$^{-1}$ MC within this distance interval.

The remaining components are categorized as background MCs. In the 5.0--8.03~kpc slice, the extinction structures exhibit a fragmented morphology, a result of the cumulative foreground extinction and the inherent increase in 3D modeling uncertainties at greater distances. Notably, the obscuration features produced by the dense regions of the $[-52, -40]$~km~s$^{-1}$ MC remain clearly discernible in the southern portion of this slice. This persistent shadowing effect at larger distances provides independent verification for the foreground placement of the target MC. In the northern part, the identified high-extinction structures correlate with regions of weaker CO emission from the $[-76, -56]$~km~s$^{-1}$ MC, effectively anchoring this component within the 5.0--8.03~kpc distance interval. Regarding the $[-120, -100]$~km~s$^{-1}$ and $[-96, -82]$~km~s$^{-1}$ MCs, no matching extinction features or significant jumps were detected within 10~kpc, suggesting their true distances exceed the model's detection limit.

Based on the FWHM of the extinction gradient peaks, we determined the precise distances for the two candidate components, as summarized in Table~\ref{tab2}. The $[-28, -18]$~km~s$^{-1}$ MC is placed at $1.09^{+0.11}_{-0.09}$~kpc, while the $[-52, -40]$~km~s$^{-1}$ MC is anchored at $2.76^{+0.37}_{-0.31}$~kpc. Additionally, the background $[-76, -56]$~km~s$^{-1}$ MC is positioned at $6.64^{+0.67}_{-0.81}$~kpc.

The study by \citet{2011A&A...526A..82E} suggested an association with the $[-27, -18]$~km~s$^{-1}$ velocity range based on an HI density-depression feature; however, this association lacks sufficient kinematic evidence of a physical SNR--MC interaction. In contrast, \citet{2025ApJ...990..213H} found that the $[-48, -43]$~km~s$^{-1}$ MC envelops the remnant in a bubble-like structure with an expansion velocity of $\sim$5~km~s$^{-1}$. This association is further supported by HI self-absorption features and the high spatial consistency between GeV $\gamma$-ray emission and the cloud morphology. Consequently, we regard the interaction proposed by \citet{2025ApJ...990..213H} as more reliable. By adopting the distance of the $[-52, -40]$~km~s$^{-1}$ MC, we assign SNR G335.2$+$0.1 a distance of $2.76^{+0.37}_{-0.31}$~kpc. This result is consistent within uncertainties with the 3.1~kpc kinematic distance estimate obtained by \citet{2025ApJ...990..213H}, effectively resolving the distance ambiguity through the integration of 3D extinction and multi-wavelength evidence.

\subsection{G351.7$+$0.8}

G351.7+0.8 is a shell-type SNR. Based on morphological analysis, \citet{2007MNRAS.378.1283T} suggested that the $[-18, -10]$~km~s$^{-1}$ HI gas is interacting with the remnant. While explicit reporting of a direct MC interaction is limited in previous literature, the kinematic evidence from HI studies provides a robust basis for identifying potential molecular counterparts. Mopra observations toward this sightline reveal an extremely complex interstellar environment, characterized by seven primary MC components with velocity ranges of: $[-116, -86]$~km~s$^{-1}$, $[-72, -60]$~km~s$^{-1}$, $[-60, -40]$~km~s$^{-1}$, $[-36, -24]$~km~s$^{-1}$, $[-22, -14]$~km~s$^{-1}$, $[-10, 2]$~km~s$^{-1}$, and $[4, 8]$~km~s$^{-1}$ (see Figures~\ref{fig 1}i and \ref{fig:A2}). Following the HI kinematic constraints by \citet{2007MNRAS.378.1283T}, the $[-22, -14]$~km~s$^{-1}$ MC is identified as the most likely candidate for physical association with the SNR.

\begin{figure}[htb!]
    \centering
    \includegraphics[width=0.9\textwidth]{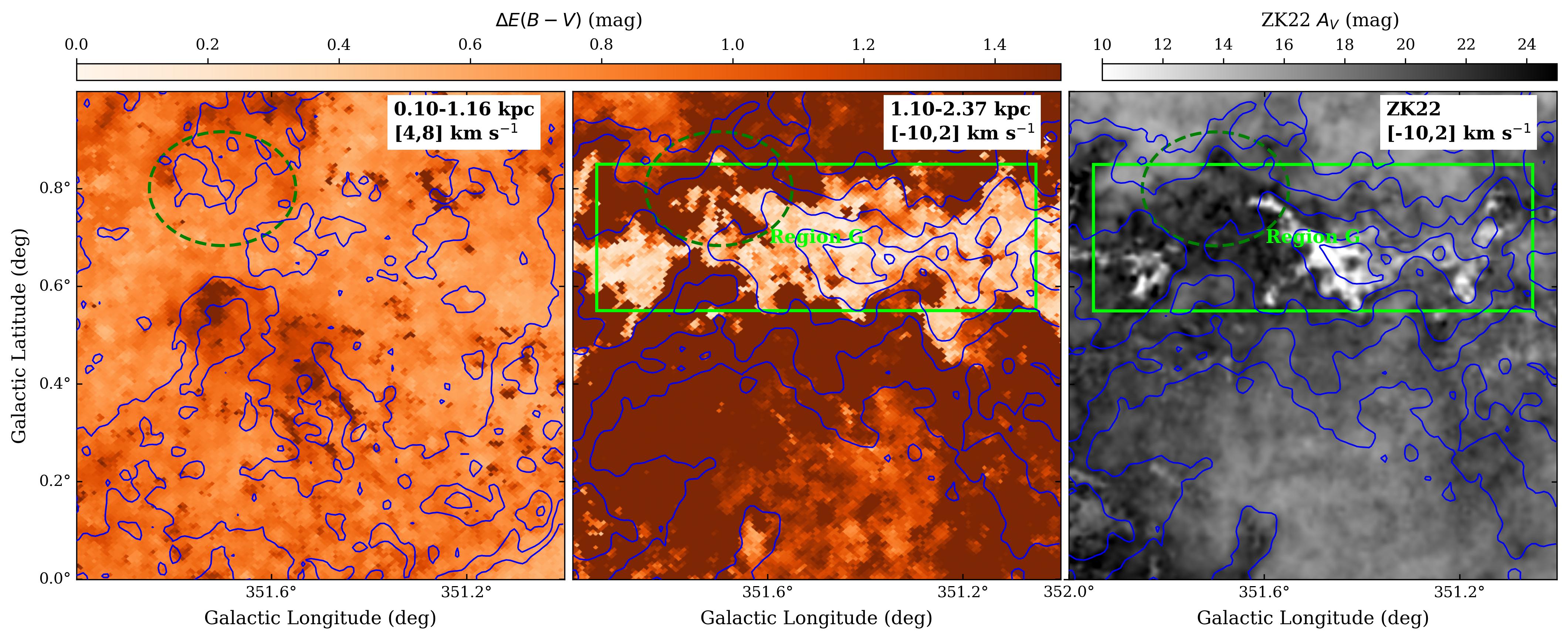}  
    \caption{Same as Figure \ref{fig 4}, but for G351.7$+$0.8. Note that the 2D extinction map in the final panel is from \protect\hyperlink{ZK22_definition}{ZK22}.}
    \label{fig 12}  
\end{figure}

We identified only two of the seven MC components that can be morphologically matched with the differential extinction maps within 10~kpc (see Figure~\ref{fig 12}). In the 0.53--1.03~kpc interval, the extinction structures exhibit a broad distribution, showing excellent morphological agreement with the $[4, 8]$~km~s$^{-1}$ MC. The second identifiable component, the $[-10, 2]$~km~s$^{-1}$ MC, is anchored within the 1.09--1.64~kpc interval. In this slice, Region~G (in the south-central part of the remnant, Figure~\ref{fig 12}) exhibits a prominent extinction-missing structure where the integrated CO intensity reaches $\sim$300~K~km~s$^{-1}$. This implies an extraordinary expected extinction of $E(B-V) \approx 21$~mag, a feature clearly corroborated by the \hyperlink{ZK22_definition}{ZK22} 2D extinction map. 

Based on the FWHM of the extinction gradient peaks, the distances for these two identifiable components are determined to be $0.52^{+0.22}_{-0.26}$~kpc (for the $[4, 8]$~km~s$^{-1}$ MC) and $1.34^{+0.14}_{-0.14}$~kpc (for the $[-10, 2]$~km~s$^{-1}$ MC). Regarding previous estimates, \citet{2007MNRAS.378.1283T} derived a kinematic distance of 13.2~kpc, whereas \citet{2020A&A...639A..72W} obtained a closer distance of $3.35 \pm 0.11$~kpc using red-clump stars. However, the extreme column density within Region~G at 1.34~kpc creates a near-impenetrable barrier that obscures almost all background stars, effectively masking the extinction signals from more distant layers. Consequently, a reliable distance for the target $[-22, -14]$~km~s$^{-1}$ MC cannot be retrieved. Given that this dense foreground cloud completely shields the target region, we adopt $1.34$~kpc as a robust lower limit for the distance to SNR G351.7+0.8.

\subsection{G353.6$-$0.7}

G353.6$-$0.7 is a typical shell-type SNR, generally considered to be physically associated with the TeV $\gamma$-ray source HESS~J1731$-$347 \citep{2008ApJ...679L..85T}. \cite{2014ApJ...788...94F} studied the HI and CO molecular gas in the velocity range \([-90, -75]\)~km~s\(^{-1}\) and suggested that the remnant may be interacting with gas in this velocity interval. \cite{2008ApJ...679L..85T} noted that the bright HII region G353.42$-$0.37 near the remnant is likely associated with G353.6$-$0.7, and, using its HI absorption spectrum, estimated the LSR velocity of the remnant to be $-16$~km~s\(^{-1}\). \cite{2018MNRAS.474..662M} compared the X-ray absorption column densities at several positions of the remnant’s shell with the total gas column density along the line of sight derived from CO and HI emission, finding convergence at a velocity of $-15$~km~s\(^{-1}\). They also detected CS $(1-0)$ emission from dense gas at this velocity, revealing a potential SNR--MC interaction. Based on Mopra data, we identified four main MC components in the region, with velocity ranges of \([-90, -70]\)~km~s\(^{-1}\), \([-60, -40]\)~km~s\(^{-1}\), \([-24, -12]\)~km~s\(^{-1}\), and \([-2, 4]\)~km~s\(^{-1}\) (see Figures~\ref{fig 1}(j) and \ref{fig:A2}). Integrating the results of \citet{2014ApJ...788...94F} and \citet{2018MNRAS.474..662M}, we adopt the \([-90, -70]\)~km~s\(^{-1}\) and \([-24, -12]\)~km~s\(^{-1}\) components as the target MCs that may be interacting with the remnant.

\begin{figure}[htb!]
    \centering
    \includegraphics[width=0.9\textwidth]{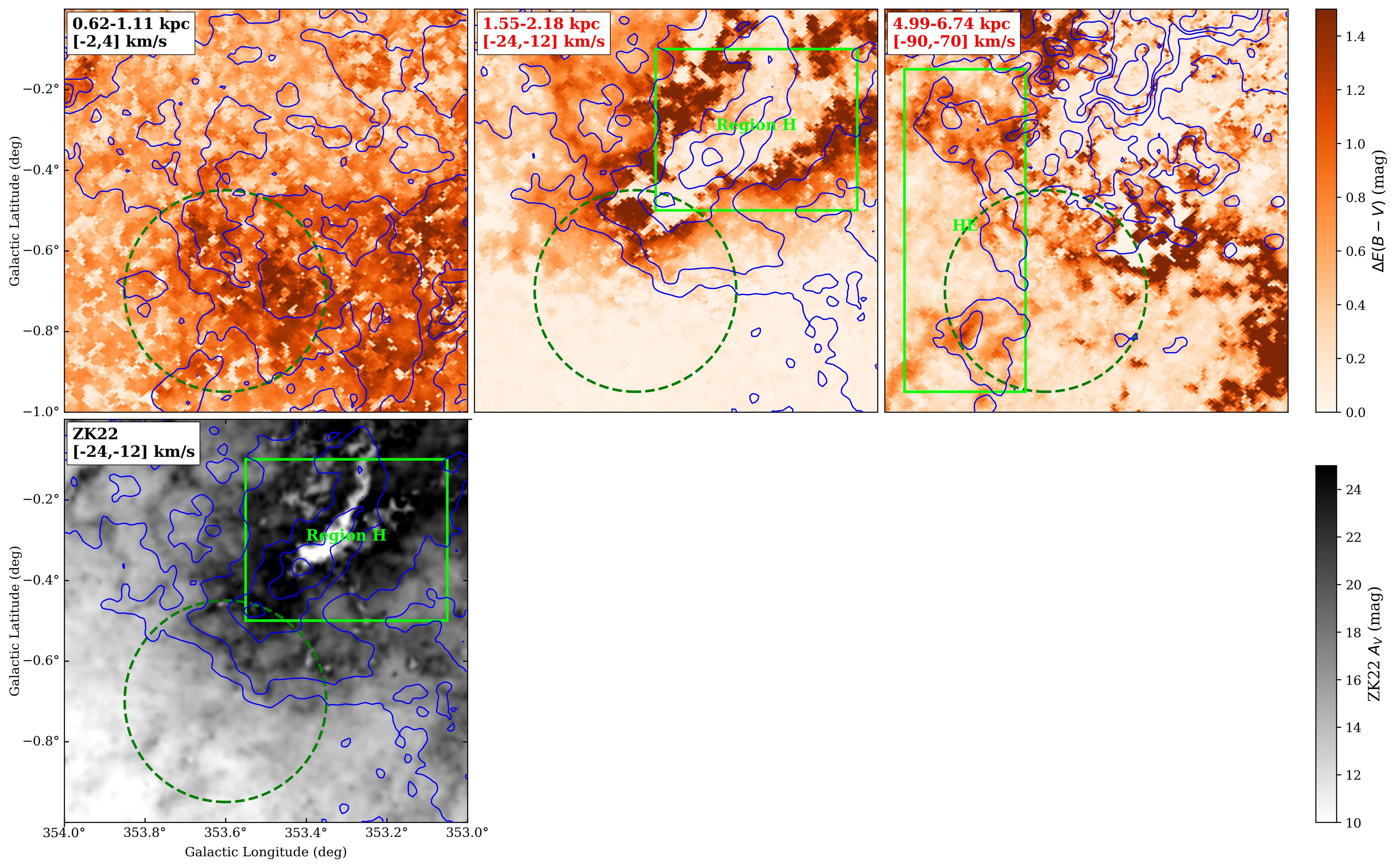}  
    \caption{Same as Figure \ref{fig 4}, but for G353.6$-$0.7. Note that the 2D extinction map in the final panel is from \protect\hyperlink{ZK22_definition}{ZK22}.}
    \label{fig 13}  
\end{figure}

Regarding the candidate components, the $[-24, -12]$~km~s$^{-1}$ MC is anchored within the 1.55--2.18~kpc interval (see Figure~\ref{fig 13}). In this slice, Region~H (Figure~\ref{fig 13}), in the northwestern part of the remnant, exhibits a prominent extinction-missing structure where the integrated CO intensity reaches $\sim$150~K~km~s$^{-1}$ ($E(B-V) \approx 10.5$~mag). This signal deficit is physically consistent with the high extinction opacity revealed in the \hyperlink{ZK22_definition}{ZK22} 2D map. Furthermore, consistent high-extinction features are clearly observed at the cloud's periphery where the CO emission is relatively weaker, showing excellent morphological agreement with the MC. For the second candidate, the $[-90, -70]$~km~s$^{-1}$ MC is positioned within the 4.99--6.74~kpc interval. Due to the intense foreground obscuration from the $[-24, -12]$~km~s$^{-1}$ MC, the extinction features in the northwestern portion of this distant slice are largely fragmented. However, within Region~HE (in the mid-western part of the remnant), the extinction features remain well-defined and exhibit strong morphological consistency with the spatial distribution of the \([-90, -70]\)~km~s\(^{-1}\) MC.

Other velocity components include the foreground $[-2, 4]$~km~s$^{-1}$ MC, which aligns well with structures in the 0.62--1.11~kpc slice. No extinction features were detected for the $[-60, -40]$~km~s$^{-1}$ MC, likely due to its location beyond 10~kpc or foreground shielding. 

Based on the FWHM of the extinction gradient peaks, the distances for the identified components are determined as follows: the foreground $[-2, 4]$~km~s$^{-1}$ MC is placed at $0.85^{+0.15}_{-0.14}$~kpc, while the two candidate components, the $[-24, -12]$~km~s$^{-1}$ and $[-90, -70]$~km~s$^{-1}$ MCs, are anchored at $1.81^{+0.18}_{-0.14}$~kpc and $6.13^{+0.22}_{-0.21}$~kpc, respectively.

Compared to the association with the $[-90, -70]$~km~s$^{-1}$ MC proposed by \citet{2014ApJ...788...94F} based on morphological similarities, we favor the interaction with the $\sim$$-15$~km~s$^{-1}$ MC, supported by the analysis of X-ray absorption and gas column densities \citep{2018MNRAS.474..662M}. This preference is rooted in the clear convergence exhibited between the X-ray absorption and gas column densities at $-15$~km~s$^{-1}$, which provides a more robust physical constraint. Consequently, we determine the distance to SNR G353.6$-$0.7 to be $1.81^{+0.18}_{-0.14}$~kpc, consistent with our measured distance for the $[-24, -12]$~km~s$^{-1}$ MC. Our result is closer than previous kinematic estimates ($\sim$3.2~kpc; \citealp{2008ApJ...679L..85T, 2018MNRAS.474..662M}) and the range derived from X-ray studies (3.2--4.0~kpc; \citealp{2017A&A...608A..23D}). We attribute this significant discrepancy to peculiar motions toward this sightline and the extreme column density of the $[-24, -12]$~km~s$^{-1}$ MC, which likely compromises the accuracy of traditional kinematic models and gas-to-dust ratio assumptions in this specific direction.

\section{SUMMARY} \label{sec:SUMMARY}

In this study, we have systematically determined the distances to ten SNRs in the southern Galactic plane by integrating Mopra CO survey data with the DECaPS 3D extinction map. 
Our method follows the extinction-slicing approach originally developed by \citet{2026arXiv260320881Z}, which anchors distances through spatial-morphological matching between MCs and 3D differential extinction structures. 
Building upon this framework, we present two key innovations: (i) the first application of this method to a sample of southern-hemisphere SNRs, using high-resolution Mopra CO observations; and (ii) the incorporation of the \hyperlink{CSFD_def}{CSFD} or \hyperlink{ZK22_definition}{ZK22} 2D extinction maps as auxiliary diagnostics in regions of extreme extinction, where the 3D map suffers from a deficit of background stars. 
This combined use of 3D slicing and 2D verification effectively mitigates the extinction-missing bias caused by ultra-dense cloud cores, thereby enhancing the reliability of distance assignments in complex interstellar environments.

Applying this method, we have derived precise distances for nine SNRs and established a robust lower limit for one. 
The results are: 
G290.1$-$0.8 at $7.32^{+0.60}_{-0.47}~\mathrm{kpc}$, 
G292.2$-$0.5 at $10.85^{+0.43}_{-0.68}~\mathrm{kpc}$, 
G296.1$-$0.5 at $4.59^{+0.18}_{-0.19}~\mathrm{kpc}$, 
G296.8$-$0.3 at $8.74^{+0.40}_{-0.29}~\mathrm{kpc}$, 
G298.6$-$0.0 at $6.50\pm0.21~\mathrm{kpc}$, 
G312.4$-$0.4 at $3.60^{+0.19}_{-0.23}~\mathrm{kpc}$, 
G332.4$-$0.4 at $2.66^{+0.23}_{-0.15}~\mathrm{kpc}$, 
G335.2$+$0.1 at $2.76^{+0.37}_{-0.31}~\mathrm{kpc}$, and 
G353.6$-$0.7 at $1.81^{+0.18}_{-0.14}~\mathrm{kpc}$. 
For G351.7$+$0.8, an extremely dense foreground cloud ($E(B-V)\approx21~\mathrm{mag}$) completely obscures background stars, preventing a direct distance measurement; we therefore adopt $1.34~\mathrm{kpc}$ as a robust lower limit. 
These distances span a wide range from $1.8~\mathrm{kpc}$ to nearly $11~\mathrm{kpc}$, demonstrating the versatility of our method across different Galactic environments.

Our results show excellent agreement with prior kinematic and X-ray absorption estimates in several cases (e.g., G290.1$-$0.8, G296.8$-$0.3, G312.4$-$0.4), while revealing significant discrepancies for others (e.g., G296.1$-$0.5, G298.6$-$0.0, G332.4$-$0.4). 
We attribute these deviations primarily to non-circular gas motions and the limitations of traditional rotation-curve models in the inner Galaxy. 
Notably, for G290.1$-$0.8 we demonstrate that a previously reported extinction distance of $\sim$3.4~kpc \citep{2025ApJ...988..176C} actually originates from a foreground MC, illustrating the power of our spatial-morphological approach to disentangle line-of-sight contamination.

Despite the progress, several limitations remain. 
The identification of physical SNR-MC associations still largely relies on prior multi-wavelength evidence; direct dynamical indicators such as CO line broadening or OH maser emission are not available for most of our targets. 
Moreover, the probing depth of current 3D extinction maps is ultimately limited by the background star density, which remains insufficient to penetrate the most opaque regions of the southern Galactic plane (e.g., the foreground cloud toward G351.7$+$0.8). 
Looking forward, the forthcoming Gaia DR4 is expected to greatly improve the spatial resolution and depth of 3D dust maps. 
Combined with more sensitive multi-tracer molecular observations (e.g., CS, HCO$^{+}$) and systematic maser surveys, our method can be extended to a larger sample of SNRs, providing crucial observational constraints on the distribution of core-collapse supernovae and their feedback effects on the ISM.

\begin{acknowledgments}
We are grateful to Drs. Shu Wang and Jian Gao for their helpful discussions and suggestions. This work is supported by the National Natural Science Foundation of China (NSFC) project Nos. 12133002 and 12403026. H.Z. acknowledges financial support by the Chilean Government-ESO Joint Committee (Comit\'e Mixto ESO-Chile, No. annlang23003-es-cl). We thank the Mopra Southern Galactic Plane CO Survey team for making their data publicly available. The Mopra radio telescope is part of the Australia Telescope National Facility, which is funded by the Commonwealth of Australia for operation as a National Facility managed by CSIRO. We acknowledge the use of data from the DECaPS 3D dust extinction map, and thank the teams behind the \hyperlink{ZK22_definition}{ZK22} and \hyperlink{CSFD_def}{CSFD} 2D extinction maps for providing their valuable data products. The radio continuum data from the EMU and the SUMSS were essential for identifying SNR morphologies. We also thank D. A. Green, G. Ferrand, and S. Safi-Harb for maintaining the Galactic SNR catalogue.

\end{acknowledgments}

\vspace{5mm}
\facilities{Mopra-22m}

\software{Astropy \citep{AstropyCollaboration2013A&A},
          APLpy \citep{Robitaille2012ascl.soft},
          dustmaps \citep{Green2018JOSS},
          Spectral-cube \citep{Ginsburg2015ASPC}
          }

\bibliography{sample701}{}
\bibliographystyle{aasjournalv7}

\begin{table*}[t!] 
\centering
\caption{SNR information and proposed velocity ranges for the associations}
\label{tab1}

\begin{tabular}{l c c c c}
\hline\hline
Name & Size & $V_{\rm LSR}(\text{literature})$ & $V_{\rm LSR}(\text{this work})$ & References \\
     & (arcmin) & (km s$^{-1}$) & (km s$^{-1}$) &  \\
\hline
G290.1$-$0.8 & 19$\times$14 & 12 (H$\alpha$), [7, 23] (CO), 7 (HI) & [8, 26] & 1, 2, 3 \\
G292.2$-$0.5 & 20$\times$15 & [16, 28] (HI), [20, 30] (CO) & [22, 34] & 4, 5 \\
G296.1$-$0.5 & 37$\times$25 & $-$35 (H$\alpha$, H$\beta$), [$-$30, $-$15] (CO) & [$-$34, $-$24] & 6, 7 \\
G296.8$-$0.3 & 20$\times$14 & [15, 30] (HI) & [8, 30] & 8 \\
G298.6$-$0.0 & 12$\times$9 & [19, 34] (CO) & [14, 34] & 9 \\
G312.4$-$0.4 & 38 & $-$49 (H$_{\text{2}}$CO), [20, 40] (HI), $-$50 (HI) & [$-$52, $-$44] & 10, 11, 12 \\
G332.4$-$0.4 & 10 & $-$48 (HCO$^+$, CO) & [$-$52, $-$40] & 13 \\
G335.2$+$0.1 & 21 & [$-$27, $-$18] (HI, CO), [$-$48,$-$43] (CO) & [$-$28, $-$18], [$-$52, $-$40] & 14, 15 \\
G351.7$+$0.8 & 18$\times$14 & [$-$18, $-$10] (HI) & [$-$22, $-$10] & 16 \\
G353.6$-$0.7 & 30 & [$-$90, $-$75] (HI, CO), -18 (HI), -15 (HI, CO) & [$-$90, $-$70], [$-$24, $-$14] & 17, 18, 19 \\
\hline
\end{tabular}

\vspace{2mm} 

\begin{minipage}{\textwidth}
    \small
    \textbf{Notes.} $V_{\rm LSR}(\text{literature})$ represents either a single central velocity or a velocity range, depending on the data reported in the cited literature. The specific tracers used for velocity determination (e.g., CO, HI, H$\alpha$) are provided in parentheses. $V_{\rm LSR}(\text{this work})$ denotes the velocity integration intervals of the candidate MC components; these components were initially identified based on literature and subsequently refined with Mopra CO data to optimize the morphological boundaries and velocity ranges.
    
    \textbf{References:} (1)\cite{1996A&A...315..243R}, (2)\cite{2005SerAJ.170...47F}, (3)\cite{2006MNRAS.369..416R}, (4)\cite{2004MNRAS.352.1405C}, (5)\cite{2019PASA...36...14V}, (6)\cite{1977ApJ...216..291S}, (7)\cite{2022ApJ...933..101T}, (8)\cite{1998MNRAS.296..813G}, (9)\cite{2023PASJ...75..384Y}, (10)\cite{1985MNRAS.216..753C}, (11)\cite{2003MNRAS.339.1048D}, (12)\cite{2022ApJ...940...63R}, (13)\cite{2006PASA...23...69P}, (14)\cite{2011A&A...526A..82E}, (15)\cite{2025ApJ...990..213H}, (16)\cite{2007MNRAS.378.1283T}, (17)\cite{2014ApJ...788...94F}, (18)\cite{2008ApJ...679L..85T}, (19)\cite{2018MNRAS.474..662M}.
\end{minipage}

\end{table*}


\begin{table*}[ht!]
\centering
\caption{MC components and corresponding distances}
\label{tab2}
\small

\begin{tabular}{l l l l}
\hline\hline
Name & Velocity range & Distance & Representative Sightline\\
     & (km s$^{-1}$) & kpc & ($l, b$) [deg]\\
\hline
G290.1$-$0.8 & [$-$22, $-$14] & 3.36$^{+0.20}_{-0.20}$ & (289.58, $-$0.71)\\
             & [$-$12, 2]     & 4.33$^{+0.24}_{-0.15}$ & (290.20, $-$0.52)\\
             & \textbf{[8, 26]} & \textbf{7.32$^{+0.60}_{-0.47}$} & (290.09, $-$0.99)\\
             & [26, 36]      & 9.33$^{+0.55}_{-0.43}$ & (289.92, $-$0.57)\\
\hline
G292.2$-$0.5 & [$-$30, $-$16] & 2.45$^{+0.19}_{-0.20}$ & (291.80, $-$0.88) \\
             & [$-$16, 6]     & 1.35$^{+0.12}_{-0.17}$ & (292.45, $-$0.48)\\
             & \textbf{[20, 36]} & \textbf{10.85$^{+0.43}_{-0.68}$} & (291.65, $-$0.14) \\
\hline
G296.1$-$0.5 & [$-$40, $-$34] & 2.78$^{+0.15}_{-0.11}$ & (296.00, $-$0.64) \\
             & \textbf{[$-$34, $-$24]} & \textbf{4.59$^{+0.18}_{-0.19}$} & (295.87, $-$0.17) \\
             & [$-$18, $-$14] & 0.80$^{+0.14}_{-0.16}$ & (295.75, $-$0.37) \\
             & [$-$8, 0]      & 6.83$^{+0.68}_{-0.52}$ & (296.48, $-$0.69)\\
             & [10, 28]       & 8.96$^{+0.51}_{-0.49}$ & (296.47, $-$0.90)\\
\hline
G296.8$-$0.3 & [$-$38, $-$24] & 2.91$^{+0.15}_{-0.11}$ & (296.90, $-$0.88)\\
             & [$-$12, 2]     & 7.12$^{+0.50}_{-0.62}$ & (297.47, $-$0.56)\\
             & \textbf{[8, 36]} & \textbf{8.74$^{+0.40}_{-0.29}$} & (297.21, $-$0.62)\\
\hline
G298.6$-$0.0 & [$-$42, $-$28] & 2.76$^{+0.07}_{-0.09}$ & (298.65, 0.22)\\
             & [$-$28, $-$24] & 1.94$^{+0.26}_{-0.14}$ & (298.90, 0.45)\\
             & [$-$22, $-$8]  & 5.83$^{+0.24}_{-0.34}$ & (298.02, 0.30)\\
             & \textbf{[14, 34]} & \textbf{6.50$^{+0.21}_{-0.21}$} & (298.19, $-$0.49)\\
\hline
G312.4$-$0.4 & \textbf{[$-$52, $-$44]} & \textbf{3.60$^{+0.19}_{-0.23}$} & (312.18, $-$0.66)\\
             & [$-$38, $-$24] & 1.96$^{+0.10}_{-0.16}$ & (312.73, $-$0.50)\\
\hline
G332.4$-$0.4 & \textbf{[$-$52, $-$44]} & \textbf{2.66$^{+0.23}_{-0.15}$} & (332.35, $-$0.41)\\
\hline
G335.2$+$0.1 & [$-$76, $-$56] & 6.64$^{+0.67}_{-0.81}$ & (334.55, 0.40)\\
             & \textbf{[$-$52, $-$40]} & \textbf{2.76$^{+0.37}_{-0.31}$} & (335.20, 0.26)\\
             & [$-$28, $-$18] & 1.09$^{+0.11}_{-0.09}$ & (335.16, $-$0.10)\\
\hline
G351.7$+$0.8 & [$-$10, 2] & 1.34$^{+0.14}_{-0.14}$ & (351.40, 0.55)\\
             & [4, 8]     & 0.52$^{+0.22}_{-0.26}$ & (351.72, 0.55)\\
\hline
G353.6$-$0.7 & [$-$90, $-$70] & 6.13$^{+0.22}_{-0.21}$ & (353.67, $-$0.31)\\
             & \textbf{[$-$24, $-$12]} & \textbf{1.81$^{+0.18}_{-0.14}$} & (353.58, $-$0.37)\\
             & [$-$2, 4]   & 0.85$^{+0.15}_{-0.14}$ & (353.45, $-$0.75)\\
\hline
\end{tabular}

\begin{minipage}{\textwidth}
    \small
    \textbf{Notes.} Velocity range: Refers to the velocity range for all MC components along the line of sight for which a reliable distance could be determined in this study.
    Distance: Corresponds to the median distance associated with the gradient peak position in the extinction-distance profile extracted along the representative sightline. The uncertainty is defined by the FWHM of these gradient peaks.
    Representative Sightline: The specific Galactic coordinates ($l, b$) used for distance determination. These sightlines are selected by identifying regions with prominent extinction gradients within 1~kpc differential extinction maps. Notably, these positions do not always coincide with the peaks of CO integrated intensity, as extremely dense regions may suffer from the extinction-missing effect in 3D maps due to the depletion of background star tracers.
    Bold entries: Highlight the candidate molecular components most likely associated with the target SNRs and their distance estimates; non-bold entries represent the distance estimates for other identifiable foreground or background components along the same line of sight. For sources with two candidate components along the line of sight (i.e., G335.2$+$0.1 and G353.6$-$0.7), only the single MC component demonstrating the most robust physical evidence of interaction is highlighted in bold after detailed analysis.
\end{minipage}

\end{table*}


\begin{table*}[t!] 
\centering
\caption{The distances to SNRs.}
\label{tab3}

\begin{tabular}{l c c c c}
\hline\hline
Name & Dist (this work) & Dist (literature) & Methods & References \\
     & kpc & kpc   \\
\hline
G290.1$-$0.8 & 7.32$^{+0.60}_{-0.47}$ & 3.4--4.0, 7.0, 7.0--8.0, 7.0$\pm$1.0, 3.377 & k, k, k, k, e & 1, 2, 3, 4, 5 \\
G292.2$-$0.5 & 10.85$^{+0.43}_{-0.68}$ & 8.4$\pm$0.4, 3.6--6.3, 8.6--9.7, 2.422 & k, k, k, e & 6, 7, 8, 5 \\
G296.1$-$0.5 & 4.59$^{+0.18}_{-0.19}$ & 3$\pm$1, 4.0, 7.7, 6.6, 4.9, 3.8$\pm$0.5, 2.1$^{+0.9}_{-0.6}$ & e, k, $\Sigma$, $\Sigma$, $\Sigma$, R, k & 9, 9, 10, 11, 12, 13, 14 \\
G296.8$-$0.3 & 8.74$^{+0.40}_{-0.29}$ & 9.6$\pm$0.6, $>9.0$, 3.294 & k, X, e & 15, 16, 5 \\
G298.6$-$0.0 & 6.50$^{+0.21}_{-0.21}$ & 10.1$\pm$0.5 & k & 17 \\
G312.4$-$0.4 & 3.60$^{+0.19}_{-0.23}$ & $>3.8$, $>6.0$, 3.5$\pm$0.5, 4.41$\pm$0.5, $<4.0$  & k, k, k, R, o & 18, 19, 20, 13, 21 \\
G332.4$-$0.4 & 2.66$^{+0.23}_{-0.15}$ & 3.3, 3.1, 3.3, 3.0$\pm$0.3, 2.504 & k, k, k, R, e  & 22, 23, 24, 25, 5 \\
G335.2$+$0.1 & 2.76$^{+0.37}_{-0.31}$ & 1.8, 3.1 & k, k & 26, 27 \\
G351.7$+$0.8 & $>1.34$ & 13.2, 3.35$\pm$0.11 & k, R & 28, 13 \\
G353.6$-$0.7 & 1.81$^{+0.18}_{-0.14}$ & 5.2--6.0, 3.2$\pm$0.8, 3.2--4.0, 3.2 & k, k, X, k & 29, 30, 31, 32 \\ 
\hline
\end{tabular}

\vspace{2mm} 

\begin{minipage}{\textwidth}
    \small
    \textbf{Method codes:} k = kinematics, e = extinction, R = Red clump stars, X = X-ray absorption, $\Sigma$ = $\Sigma$-D relation, o = other methods.
    
    \textbf{References:} (1)\cite{1973ApL....15...61D}
    (2)\cite{1996A&A...315..243R}, (3)\cite{2005SerAJ.170...47F}, (4)\cite{2006MNRAS.369..416R}, (5)\cite{2025ApJ...988..176C},
    (6)\cite{2004MNRAS.352.1405C}, (7)\cite{2005ApJ...619..856G}, (8)\cite{2019PASA...36...14V}, 
    (9)\cite{1977MNRAS.181..541L}, (10)\cite{1983ApJ...271L..55C}, (11)\cite{1998ApJ...504..761C}, (12)\cite{1973Natur.246...28C}, (13)\cite{2020A&A...639A..72W}, (14)\cite{2022ApJ...933..101T}, 
    (15)\cite{1998MNRAS.296..813G}, (16)\cite{2012Ap&SS.337..573S}, 
    (17)\cite{2023PASJ...75..384Y}, 
    (18)\cite{1985MNRAS.216..753C}, (19)\cite{2003MNRAS.339.1048D}, (20)\cite{2022ApJ...940...63R}, (21)\cite{2023ApJ...959...97C}, 
    (22)\cite{1975A&A....45..239C}, (23)\cite{2004PASA...21...82R}, (24)\cite{2006PASA...23...69P}, 
    (25)\cite{2019RAA....19...92S}, 
    (26)\cite{2011A&A...526A..82E}, (27)\cite{2025ApJ...990..213H}, 
    (28)\cite{2007MNRAS.378.1283T}, 
    (29)\cite{2014ApJ...788...94F}, (30)\cite{2008ApJ...679L..85T}, (31)\cite{2017A&A...608A..23D}, (32)\cite{2018MNRAS.474..662M}.
\end{minipage}

\end{table*}

\clearpage 

\renewcommand{\thefigure}{A\arabic{figure}} 
\renewcommand{\thetable}{A\arabic{table}}  
\setcounter{figure}{0} 
\setcounter{table}{0}  

\appendix
\section{Appendix}
\label{sec:appendix}

\begin{figure*}[htb!]
    \centering
    \includegraphics[width=1.0\textwidth]{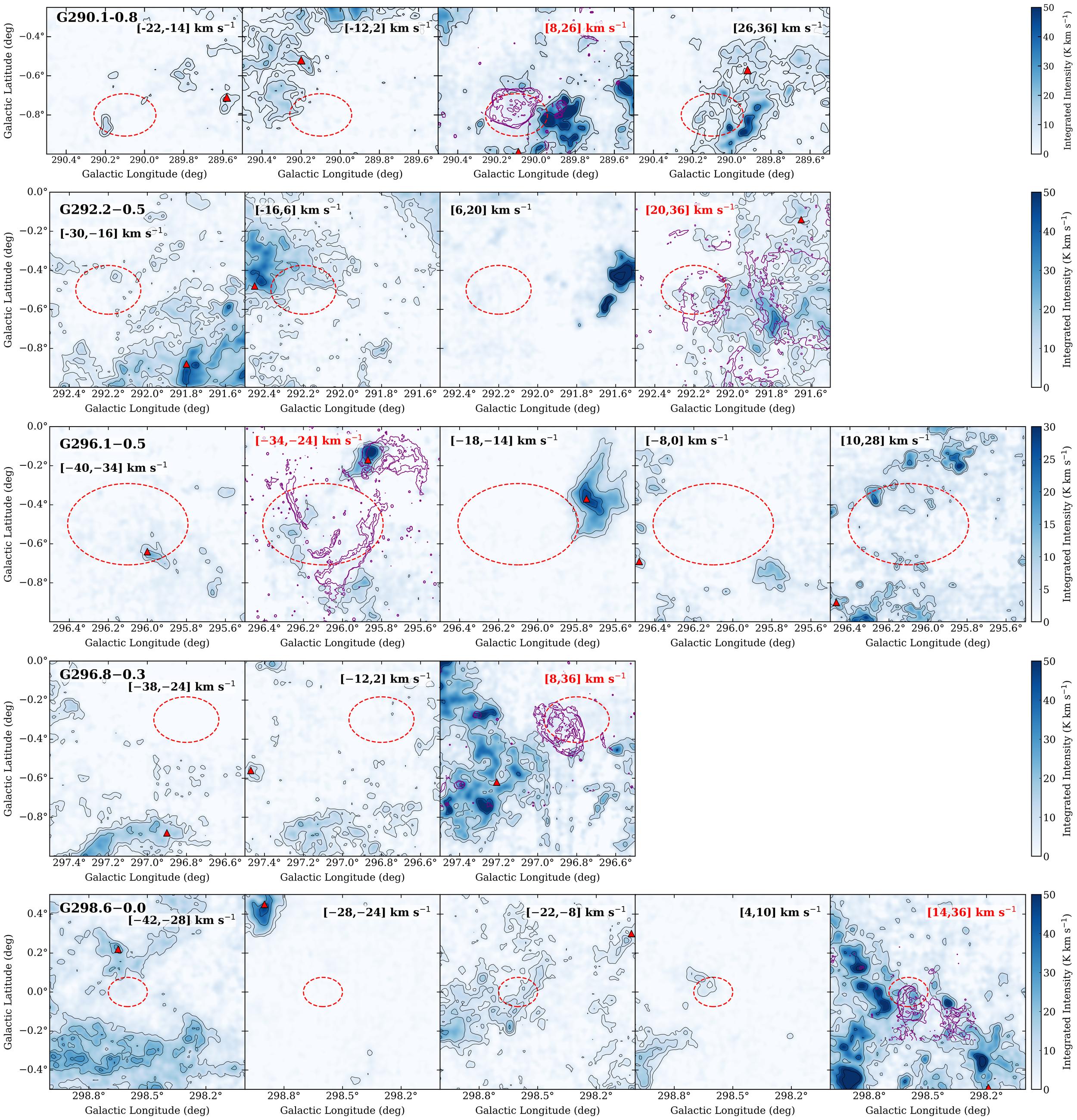}
    \caption{Velocity-integrated intensity maps of CO for the SNR regions, including G290.1$-$0.8, G292.2$-$0.5, G296.1$-$0.5, G296.8$-$0.3, and G298.6$-$0.0. The specific integrated velocity ranges are indicated in each sub-panel, with red labels marking the molecular components associated with each SNR. Black contours represent the CO integrated intensity. Purple contours indicate the radio continuum emission, tracing both the SNRs and other neighboring radio sources. Red dashed ellipses outline the approximate boundaries of the SNRs. Red triangles mark the representative CO emission positions selected for extracting the extinction--distance profiles shown in Figure~\ref{fig 3}, as well as the sightlines detailed in Table~\ref{tab2}.}
    \label{fig:A1}
\end{figure*}

\begin{figure*}[ht!]
    \centering
    \includegraphics[width=1.0\textwidth]{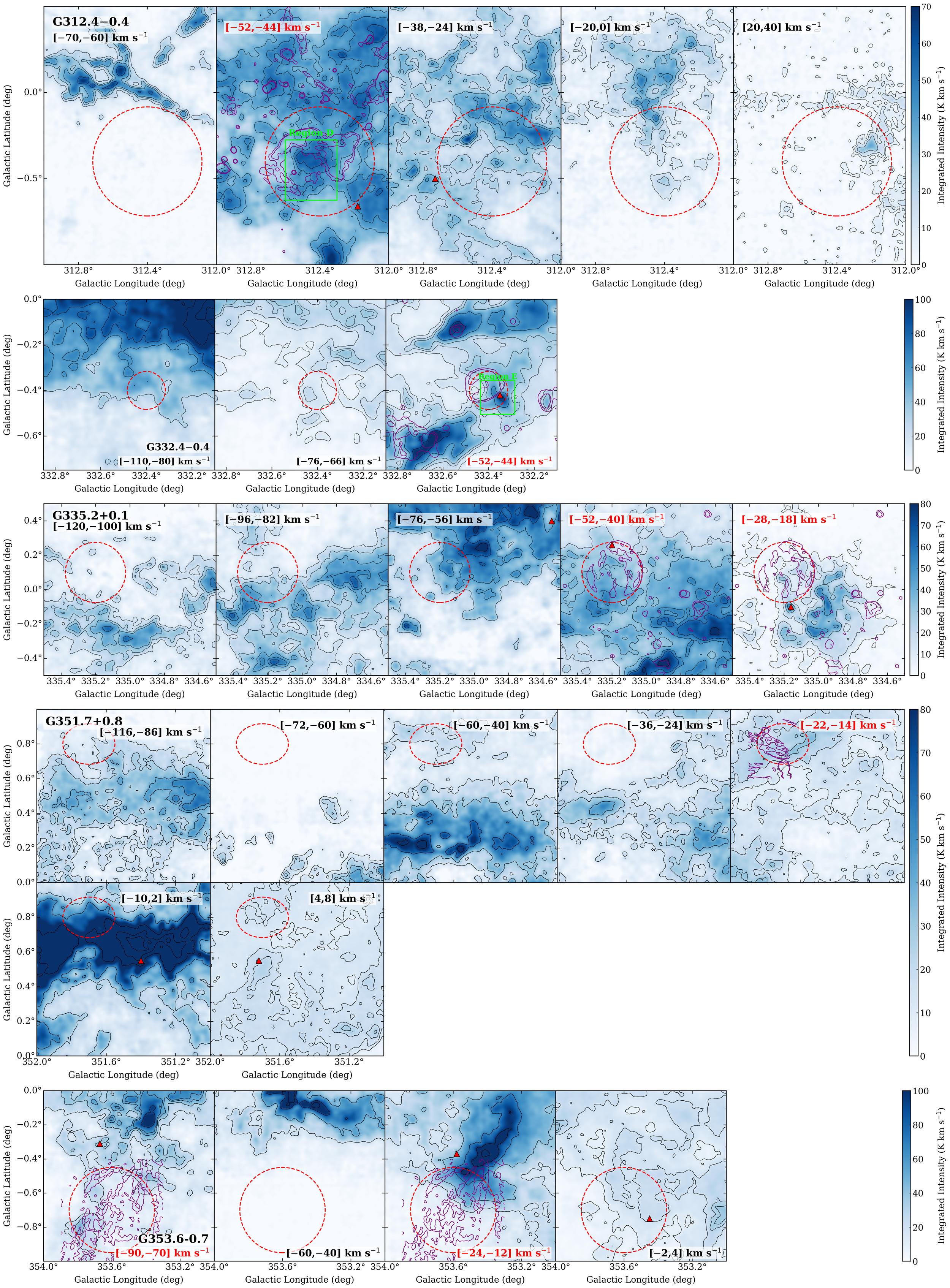}
    \caption{The same as Figure~\ref{fig:A1}, but for G312.4$-$0.4 G332.4$-$0.4, G335.2$+$0.1, G351.7$+$0.8, and G353.6$-$0.7.}
    \label{fig:A2}
\end{figure*}

\begin{figure*}[ht!]
    \centering
    \includegraphics[width=1.0\textwidth]{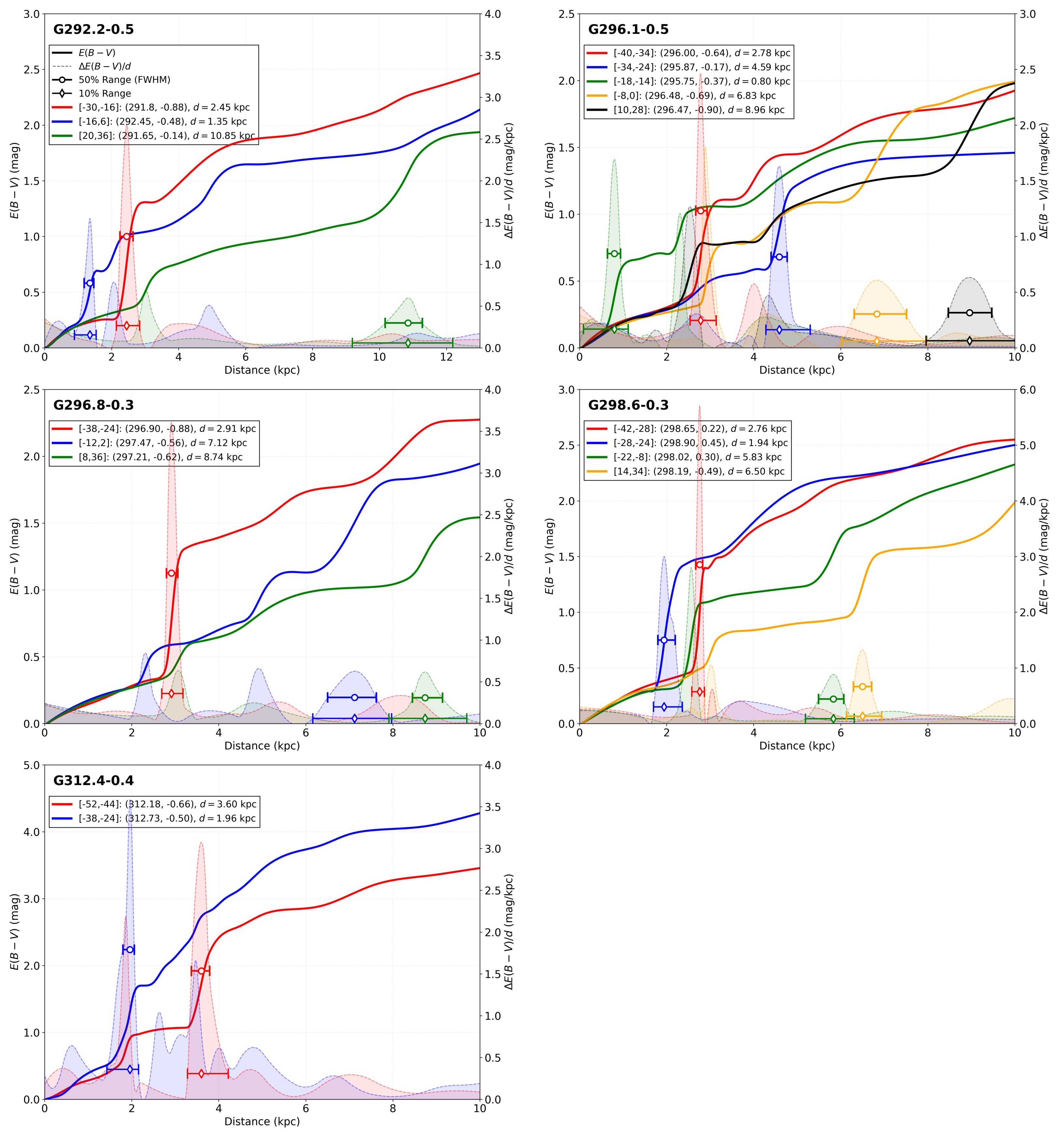}
    \caption{The same as Figure~\ref{fig 3}, but for G292.2$-$0.5, G296.1$-$0.5, G296.8$-$0.3, G298.6$-$0.0, and G312.4$-$0.4.}
    \label{fig:A3}  
\end{figure*}

\begin{figure*}[ht!]
    \centering
    \includegraphics[width=1.0\textwidth]{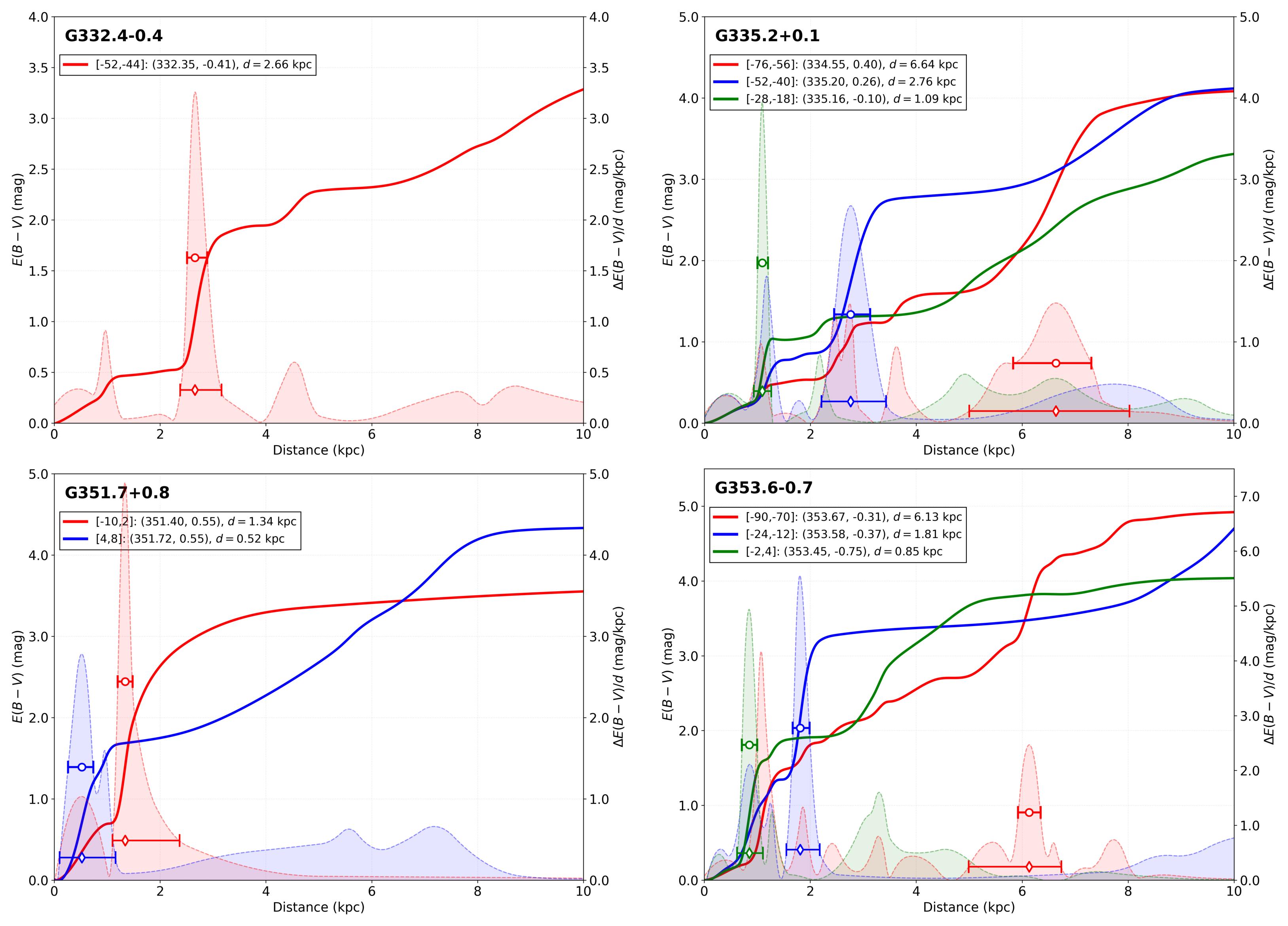}
    \caption{The same as Figure~\ref{fig 3}, but for G332.4$-$0.4, G335.2$+$0.1, G351.7$+$0.8 and G353.6$-$0.7.}
    \label{fig:A4}  
\end{figure*}

\end{CJK*}
\end{document}